\pdfoutput=1
\documentclass[11pt,twoside,a4paper,cmspaper,final,collab]{cms-tdr}
\def\svnVersion{f6cec68}\def\svnDate{2023/07/13}\def\cmsCernNoTag{CERN-EP-2022-144}\def\cmsCernDate{\today}\def\cmsMessage{Published in the European Physical Journal C as \href{http://dx.doi.org/10.1140/epjc/s10052-023-11753-y}{\doi{10.1140/epjc/s10052-023-11753-y}.}}
\begin{document}\cmsNoteHeader{SMP-21-006}

\newlength\cmsFigWidth
\ifthenelse{\boolean{cms@external}}{\setlength\cmsFigWidth{0.85\columnwidth}}{\setlength\cmsFigWidth{0.4\textwidth}}

\newlength\cmsTabSkip\setlength{\cmsTabSkip}{1ex}
\newenvironment{tolerant}[1]{\par\tolerance=#1\relax}{\par}%

\newcommand{\PYTHIAE}{{\PYTHIA\!8}\xspace}
\newcommand{\PYTHIAsix}{{\PYTHIA\!6}\xspace}
\newcommand{\MG}{\textsc{MG5\_aMC}}
\newcommand{\aMCcas}{{\textsc MG5\_aMC+CA3}\xspace}
\newcommand{\aMCPy}{{\textsc MG5\_aMC+Py8}\xspace}
\newcommand{\PBM}{PB}
\newcommand{\MADGRAPHLO}{\textsc{MadGraph+Py8}\xspace}
\newcommand{\MADGRAPHCAS}{\textsc{MadGraph+CA3}\xspace}
\newcommand{\cascade}{\textsc{Cascade3}\xspace}
\newcommand{\dphiOneTwo}{\ensuremath{\Delta\phi_{1,2}}\xspace}
\newcommand{\alpsmz}{\ensuremath{\alpS(m_{\PZ})}\xspace}
\newcommand{\ptmax}{\ensuremath{p_{\mathrm{T1}}}\xspace}
\newcommand{\ptsec}{\ensuremath{p_{\mathrm{T2}}}\xspace}
\newcommand{\ptthird}{\ensuremath{p_{\mathrm{T3}}}\xspace}
\newcommand{\ptfourth}{\ensuremath{p_{\mathrm{T4}}}\xspace}
\newcommand{\Njets}{\ensuremath{N_{\text{jets}}}\xspace}
\newcommand{\ptHLT}{\ensuremath{p_{\mathrm{T}}^{\mathrm{HLT}}}\xspace}

\cmsNoteHeader{SMP-21-006} 

\title{Measurements of jet multiplicity and jet transverse momentum in multijet events in proton-proton collisions at \texorpdfstring{$\sqrt{s}=13\TeV$}{sqrt(s)=13 TeV}}
\titlerunning{Jet multiplicity and transverse momentum at \texorpdfstring{$\sqrt{s}=13\TeV$}{sqrt(s)=13 TeV}}
\date{\today}

\abstract{Multijet events at large transverse momentum (\pt) are measured at $\sqrt{s}=13\TeV$ using data recorded with the CMS detector at the LHC, corresponding to an integrated luminosity of $36.3\fbinv$. The multiplicity of jets with $\pt>50\GeV$ that are produced in association with a high-\pt dijet system is measured in various ranges of the \pt of the jet with the highest transverse momentum and as a function of the azimuthal angle difference \dphiOneTwo between the two highest \pt jets in the dijet system. The differential production cross sections are measured as a function of the transverse momenta of the four highest \pt jets. The measurements are compared with leading and next-to-leading order matrix element calculations supplemented with simulations of parton shower, hadronization, and multiparton interactions. In addition, the measurements are compared with next-to-leading order matrix element calculations combined with transverse-momentum dependent parton densities and transverse-momentum dependent parton shower.}
\hypersetup{
pdfauthor={CMS Collaboration},
pdftitle={Measurements of jet multiplicity and jet transverse momentum in multijet events at sqrt(s)=13 TeV},
pdfsubject={CMS},
pdfkeywords={CMS, QCD, anti-kT}}

\maketitle 

\section{Introduction}
The production of jets, which are reconstructed from a stream of hadrons coming from the fragmentation of energetic partons, is described by the theory of strong interactions, quantum chromodynamics (QCD). In proton-proton ($\Pp\Pp$) collisions, at leading order (LO) in the strong coupling \alpS , two colliding partons from the incident protons scatter and produce two high transverse-momentum (\pt ) partons in the final state. The jets that originate from such a process are strongly correlated in the transverse plane, and the azimuthal angle difference between them, \dphiOneTwo , should be close to $\pi$. However, higher-order corrections to the lowest order process will result in a decorrelation in the azimuthal plane, and \dphiOneTwo\ will significantly deviate from $\pi$. These corrections can be due to either hard parton radiation, calculated at the matrix element (ME) level at next-to-leading order (NLO), or softer multiple parton radiation described by parton showers. 
In a recent approach~\cite{Martinez:2018jxt}, transverse-momentum dependent (TMD) parton densities are obtained with the parton-branching method~\cite{Hautmann:2017xtx,Hautmann:2017fcj} (\PBM -TMDs). These \PBM -TMDs were combined with NLO ME calculations~\cite{Alwall:2014hca} supplemented with \PBM\ initial-state parton showers~\cite{Baranov:2021uol}, leading to predictions where the initial-state parton shower is determined by the \PBM -TMD densities without tunable parameters. In Drell-Yan production at the LHC this approach leads to a good description of the measurements~\cite{Martinez:2019mwt}, whereas other approaches need specific tunes. Therefore, it is interesting to study the \PBM\ prediction in a jet environment, especially since the \PBM -TMD initial-state shower also becomes important.
Although \dphiOneTwo\ is an inclusive observable, it is interesting for the theoretical understanding of the complete event to measure the multiplicity of jets in different regions of \dphiOneTwo\  and the transverse momenta of the first four jets. The \dphiOneTwo\ measurement is mainly sensitive to initial-state parton showers at an inclusive level, whereas the measurement of the jet multiplicity in different regions of \dphiOneTwo\ illustrates how many high-\pt\ jets contribute to the \dphiOneTwo\ decorrelation.

The azimuthal correlation in high-\pt\ dijet events was measured previously at: the Fermilab Tevatron in proton-antiproton collisions by the \DZERO\ Collaboration at $\sqrt{s}=1.96\TeV$~\cite{bib:D0_1,bib:D0_2}; and at the CERN LHC in $\Pp\Pp$ collisions by both the ATLAS Collaboration at $\sqrt{s}=7\TeV$~\cite{bib:ATLAS} and the CMS Collaboration at $\sqrt{s}=7$, $8$, and $13\TeV$~\cite{bib:CMS,bib:CMS_2,bib:CMS_2_1,bib:CMS_2_2}.

In this paper, we describe new measurements of dijet events with rapidity $\abs{y} < 2.5$ and with transverse momenta of the leading jet $\ptmax > 200\GeV$ and the subleading jet $\ptsec > 100\GeV$. The multiplicity of jets with $\pt > 50\GeV$ is measured in bins of \ptmax\ and \dphiOneTwo . The jet multiplicity in bins of \dphiOneTwo provides information on the \dphiOneTwo decorrelation. 
The cross sections for the four leading jets are measured as a function of \pt of each jet, which can give additional information on the structure of the higher-order corrections.

This paper is organized as follows; in Section~\ref{sec:detector}, a brief summary of the CMS detector and the relevant components is given. In Section~\ref{sec:theory}, the theoretical models for comparison at detector level, as well as with the final results are described.
Section~\ref{sec:analysis} gives an overview of the analysis, with the event selection, data correction, and a discussion of the uncertainties. The final results and  comparison with theoretical predictions are discussed in Section~\ref{sect:results}.

\section{The CMS detector and event reconstruction} \label{sect:CMS}\label{sec:detector}

The central feature of the CMS apparatus is a superconducting solenoid of 6\unit{m} internal diameter, providing a magnetic field of 3.8\unit{T}. Within the solenoid volume are silicon pixel and strip tracker detectors, a lead tungstate crystal electromagnetic calorimeter (ECAL), and a brass and scintillator hadron calorimeter (HCAL), each composed of a barrel part and two endcap sections. 

Events of interest are selected using a two-tiered trigger system. The first level, composed of custom hardware processors, uses information from the calorimeters and muon detectors to select events at a rate of around 100\unit{kHz} within a fixed latency of about 4\mus~\cite{Sirunyan:2020zal}. The second level, known as the high-level trigger (HLT), consists of a farm of processors running a version of the full event reconstruction software optimized for fast processing, while reduces the event rate to around 1\unit{kHz} for data storage~\cite{Khachatryan:2016bia}.

During the 2016 data-taking period, a gradual shift in the timing of the inputs of the ECAL first level trigger in the pseudorapidity region $\abs{\eta} > 2.0$, also known as ``prefiring", caused some trigger inefficiencies~\cite{Sirunyan:2020zal}. For events containing a jet with $ \pt> 100\GeV$, in the region $2.5 < \abs{\eta}< 3.0$ the efficiency loss is 10--20\%, depending on \pt , $\eta$, and the data-taking period. Correction factors were computed from data and applied to the acceptance evaluated by simulation.

The particle-flow (PF) algorithm~\cite{CMS-PRF-14-001} reconstructs and identifies each individual particle in an event, with an optimized combination of information from the various elements of the CMS detector. The energy of charged hadrons is determined from a combination of their momentum measured in the tracker and the matching ECAL and HCAL energy deposits, corrected for the response function of the calorimeters to hadronic showers. The energy of neutral hadrons is obtained from the corresponding corrected ECAL and HCAL energies. 

The candidate vertex with the largest value of summed physics-object $\pt^2$ is taken to be the primary vertex (PV) of $\Pp\Pp$ interactions as described in Section 9.4.1 of Ref.~\cite{CMS-TDR-15-02}. The physics objects are the jets, clustered using the jet finding algorithm~\cite{Cacciari:2008gp,Cacciari:2011ma} with the tracks assigned to candidate vertices as inputs, and the associated missing transverse momentum.

Jets are reconstructed from PF objects, clustered using the anti-\kt algorithm~\cite{Cacciari:2008gp, Cacciari:2011ma} with a distance parameter of $R=0.4$.
The jet momentum is determined as the vectorial sum of all particle momenta in the jet, and is found from simulation to be, typically, within 5 to 10\% of the true momentum over the entire \pt\ spectrum and detector acceptance. Additional $\Pp\Pp$ interactions within the same or nearby bunch crossings can contribute additional tracks and calorimetric energy deposits, increasing the apparent jet momentum. To mitigate this effect, tracks identified as originating from pileup vertices are discarded and an offset correction based on the average amount of transverse energy in the event per unit area is applied to correct for the remaining contributions~\cite{CMS-PAS-JME-14-001}.
Jet energy corrections are derived from simulation studies so that the average measured energy of jets becomes identical to that of particle-level jets. 
However, the selective ECAL readout leads to a bias in the jet energy scale. 
In situ measurements of the momentum balance in dijet, photon+jet, Z+jet, and multijet events are used to determine any residual differences between the jet energy scale (JES) in data and in simulation, and appropriate corrections are made~\cite{Khachatryan:2016kdb}. Additional selection criteria are applied to each jet to remove jets potentially dominated by instrumental effects or reconstruction failures. The jet energy resolution  (JER) amounts typically to 15--20\% at 30\GeV, 10\% at 100\GeV, and 5\% at 1\TeV~\cite{Khachatryan:2016kdb}.

The missing transverse momentum vector \ptvecmiss is computed as the negative vector sum of the transverse momenta of all the PF candidates in an event, and its magnitude is denoted as \ptmiss~\cite{Sirunyan:2019kia}. The \ptvecmiss is modified to include corrections to the energy scale of the reconstructed jets in the event. Anomalous high-\ptmiss events can be due to a variety of reconstruction failures, detector malfunctions, or noncollision backgrounds. Such events are rejected by event filters that are designed to identify more than 85--90\% of the spurious high-\ptmiss events with a mistagging rate less than 0.1\%~\cite{Sirunyan:2019kia}. 

A more detailed description of the CMS detector, together with a definition of the coordinate system used and the relevant kinematic variables, can be found in Ref.~\cite{Chatrchyan:2008zzk}.

\section{Theoretical predictions}
\label{sec:theory}
Theoretical predictions from Monte Carlo (MC) event generators at LO and NLO are used for comparison with measurements of the jet multiplicities as well as with the \pt\ spectra in multijet final states.

We use the following predictions at LO:
\begin{itemize}
\begin{tolerant}{2400}
\item{ \PYTHIAE}~\cite{Sjostrand:2014zea} (version 8.219) simulates LO $2\to2$ hard processes. The parton shower is generated in a phase space ordered in transverse momentum and longitudinal momentum of the emitted partons, and the colored strings are hadronized using the Lund string fragmentation model. The CUETP8M1~\cite{Khachatryan:2015pea} tune (with the parton distribution function (PDF) set NNPDF2.3LO~\cite{Ball:2011uy}) gives  the parameters for multiparton interactions (MPI).
\item{ \HERWIGpp}~\cite{Bahr:2008pv} (version 2.7.1) simulates LO $2\to2$ hard processes. The emitted partons in the parton shower follow angular ordering conditions, and the cluster fragmentation model is used to transform colored partons into observable hadrons. The CUETHppS1~\cite{Khachatryan:2015pea} tune (with the PDF set CTEQ6L1~\cite{Pumplin:2002vw})  
is applied.
\item{\MGvATNLO}~\cite{Alwall:2014hca} (version 2.3.3) event generator, labeled \MADGRAPHLO, is used in the LO mode, with up to four noncollinear high-\pt partons included in the matrix element (ME), supplemented with parton showering and MPI using \PYTHIAE with the CUETP8M1 tune and merged according to the \kt-MLM matching procedure~\cite{Alwall:2007fs} with a matching scale of $10\GeV$.
\item{\MGvATNLO}~\cite{Alwall:2014hca} (version 2.6.3) event generator, labeled \MADGRAPHCAS, is used in the LO mode to generate up to four noncollinear high-\pt partons included in the ME.
The \PBM -method describes the DGLAP evolution as a process of successive branching processes. It has been shown~\cite{Hautmann:2017xtx,Hautmann:2017fcj} that on an inclusive level, the DGLAP evolution is exactly reproduced, whereas the simulation of each branching determines the transverse momentum distribution obtained during evolution. The \PBM-TMDs are obtained from a fit to inclusive HERA deep-inelastic electron proton collision data~\cite{Martinez:2018jxt}. Since the \PBM-TMDs come from a (forward) branching evolution, the initial-state parton shower (in a backward evolution) follows directly from the \PBM-TMD distributions, and therefore no free parameters are left for the initial-state shower.
We use the TMD merging~\cite{BERMUDEZMARTINEZ2021136700} procedure for combining the TMD parton shower with the ME calculations. 
The NLO \PBM -TMD~set~2~\cite{Martinez:2018jxt} with $\alpsmz= 0.118$ is used.  
The inclusion of the transverse momentum \kt and initial-state \PBM -TMD  parton shower is performed with \cascade~\cite{Baranov:2021uol} (version 3.2.3).
The initial-state parton shower follows the \PBM -TMD distribution, and has no free parameters left.
The final-state radiation (since not constrained by TMDs) as well as hadronization is performed with \PYTHIAsix (version 6.428)~\cite{Sjostrand:2006za} with  an angular ordering veto imposed.
A merging scale value of 30\GeV is used, since it provides a smooth transition between ME and PS computations. 
MPI effects are not simulated. 
\end{tolerant}
\end{itemize}

At NLO, different theoretical predictions are used. The factorization and renormalization scales are set to half the sum over the scalar transverse momenta of all produced partons, $ 1/2 \sum H_{\mathrm{T}}$.
However, we have explicitly checked that the distributions do not change when choosing $ 1/4 \sum H_{\mathrm{T}}$ or even $ 1/6 \sum H_{\mathrm{T}}$.  
The uncertainty bands of the NLO predictions are determined from the variation of the factorization and renormalization scales by a factor of two up and down, avoiding the largest scale differences.
\begin{itemize}
\begin{tolerant}{4800}
\item {\MGvATNLO}~\cite{Alwall:2014hca} (version 2.6.3) interfaced with \PYTHIAE (version 8.306), labeled \aMCPy (jj), is used with MEs computed at NLO for the process $\Pp\Pp \to \mathrm{jj}$.
The {\textsc NNPDF 3.0 NLO} PDF set~\cite{Ball:2014uwa} is used and \alpsmz is set to 0.118. This calculation uses the CUETP8M1 tune to study the effect of MPI. 
\item  {\MGvATNLO}~\cite{Alwall:2014hca} (version 2.6.3) interfaced with \cascade (version 3.2.2)~\cite{Baranov:2021uol}, labeled \aMCcas (jj), is used with MEs computed at NLO for the process $\Pp\Pp \to \mathrm{jj}$. The \textsc{Herwig6}~\cite{Corcella:2000bw} subtraction terms are used in \MCATNLO~\cite{Frixione:2003ei,Frixione:2002ik}, since they are closest to the needs for applying \PBM -TMD parton densities, as described in Ref.~\cite{Baranov:2021uol}.
The NLO \PBM -TMD~set~2~\cite{Martinez:2018jxt} with $\alpsmz = 0.118$ is used.  
MPI are not simulated in this approach.
\item  {\MGvATNLO}~\cite{Alwall:2014hca} (version 2.6.3) interfaced with \cascade, labeled \aMCcas (jjj) NLO, is used with MEs computed at NLO for the process $\Pp\Pp \to \mathrm{jjj}$.
The same \PBM -TMD distribution and parton shower calculation procedure as for \aMCcas (jj) are used.
\end{tolerant}
\end{itemize}

In Table~\ref{tab:MC}, the Monte Carlo event generators utilized in this analysis are summarized. 
Events generated by  \PYTHIAE, \MADGRAPHLO, and \HERWIGpp  are passed through a full detector simulation based on \GEANTfour~\cite{geant}. 
The simulated events are reconstructed the same way as the observed data events.

\begin{table*}[h!]
    \centering
    \topcaption{Description of the simulated samples used in the analysis.}    
    \begin{tabular}{l l l l} 
        Generator & PDF & ME & Tune \\
        \hline
        \PYTHIAE~\cite{Sjostrand:2014zea}  & {NNPDF~2.3 (LO)}~\cite{Ball:2011uy} & LO $2\to2$ & {CUETP8M1}~\cite{Khachatryan:2015pea} \\
        \MADGRAPHLO~\cite{Alwall:2014hca}  & {NNPDF~2.3 (LO)}~\cite{Ball:2011uy} & LO $2\to 2,3,4$ & {CUETP8M1}~\cite{Khachatryan:2015pea} \\
        \MADGRAPHCAS~\cite{Alwall:2014hca}  & {\PBM -TMD set~2 (NLO)}~\cite{Martinez:2018jxt} & LO $2\to 2,3,4$ & \NA \\
        \HERWIGpp~\cite{Bahr:2008pv}  & {CTEQ6L1 (LO)}~\cite{Pumplin:2002vw}& LO $2\to2$ & {CUETHppS1}~\cite{Khachatryan:2015pea} \\
        \aMCPy (jj)&{NNPDF~3.0 (NLO)}~\cite{Ball:2014uwa} & NLO $2\to 2$ & {CUETP8M1}~\cite{Khachatryan:2015pea} \\
        \aMCcas (jj) & {\PBM -TMD set~2 (NLO)}~\cite{Martinez:2018jxt} & NLO $2\to2 $ & \NA \\
        \aMCcas (jjj) & {\PBM -TMD set~2 (NLO)}~\cite{Martinez:2018jxt} & NLO $2\to3 $ & \NA \\
    \end{tabular}

    \label{tab:MC}
\end{table*}

\section{Data analysis} \label{sec:analysis}

The measured data samples recorded in 2016, corresponding to an integrated luminosity of $36.3\fbinv$, were collected with single-jet HLTs. For each single-jet HLT at least one jet with \pt higher than the \ptHLT trigger threshold is required. All triggers, except the one with the highest \ptHLT threshold, were prescaled. We consider events only if the leading jet, reconstructed with the PF algorithm, can be matched with an HLT jet. 
In Table~\ref{table:trigger_emulation_turnons}, the integrated luminosity $\mathcal{L}$ for each trigger is shown. The trigger efficiency ($>99.5\%$) is estimated using triggers with lower \ptHLT thresholds; for the trigger with lowest \ptHLT threshold a tag-and-probe~\cite{Sirunyan:2020zal} method is used to determine the \pt threshold. 
To address the issue of trigger inefficiency resulting from prefiring (discussed in Section~\ref{sec:detector}), the simulated event samples are corrected on an event-by-event basis prior to unfolding. Maps of the prefiring probability in the region of $2.0 < \abs{\eta} < 3.0$ are utilized, taking into account the \pt and $\eta$ of the jets. The total event weight is then calculated as the product of the nonprefiring probability of all jets. For each \pt and $\eta$ bin in the prefiring maps, the uncertainty per jet is determined by taking the maximum value between 20\% of the prefiring probability and the statistical uncertainty.

\begin{table*}[h]
	\centering
	\topcaption{ The integrated luminosity for each trigger sample considered in this analysis with the \pt thresholds for HLT (PF) reconstruction.}	
    \renewcommand{\arraystretch}{1.3}
	\begin{tabular}{l c c c c c c}
		$\pt^\text{HLT}$ (\GeVns) & 140 & 200 & 260 & 320 & 400 & 450 \\
        \hline
		$\pt^\text{PF}$ (\GeVns) & 196 & 272 & 362 & 430 & 548 & 592 \\
		$\mathcal{L} (\text{pb}^{-1})$ & 24.2 & 103 & 594 & 1770 & 5190 & 36300 \\
	\end{tabular}

	\label{table:trigger_emulation_turnons}
\end{table*}

The jets are corrected using the JES correction procedure in CMS~\cite{Khachatryan:2016kdb}, and an additional smoothing procedure (described in Ref.~\cite{CMS:2021yzl}) is applied to the JES correction. 
The simulated samples are corrected to take into account the JER by spreading the \pt of the jets according to the resolution extracted from  data. Jets reconstructed in regions of the detector corresponding to noisy towers in the calorimeters are excluded from the measurement~\cite{CMS:2009gpr}.

The pileup profile used in simulation is corrected to reproduce the one in data.

\subsection{Event selection}\label{sec:selection}

Each event is required to have a reconstructed PV.  
The PV must satisfy $\abs{z_\mathrm{PV}}<24\unit{cm}$ and $\rho_\mathrm{PV}<2\cm$, where $z_\mathrm{PV}$ ($\rho_\mathrm{PV}$) is the longitudinal (radial) distance from the nominal interaction point. 

All events that contain jets clustered using the anti-\kt algorithm~\cite{Cacciari:2008gp, Cacciari:2011ma} with a distance parameter of $R=0.4$ and reconstructed with $\abs{\eta}<3.2$ and transverse momentum $\pt>20\GeV$ are preselected. 
From these events, the ones with at least a pair of jets with $\ptmax>200\GeV$, $\ptsec>100\GeV$ and $\abs{y^{{1,2}}}<2.5$ are selected (events with one of the leading two jets with $\abs{y}>2.5$ are vetoed). 
Additional jets must have $\pt > 50\GeV$ and $\abs{y} < 2.5$. In addition, jets must satisfy quality criteria based on the jet constituents, in order to reject misidentified jets~\cite{CMS-PAS-JME-16-003}. 
The selected events must have a missing transverse energy fraction smaller than 0.1 (more details in Sec.~\ref{sec:background}). The selection at particle-level includes only the \pt and $\abs{y}$ constraints on the jets.

\subsection{Observables and phase space}\label{sec:observables}
In the following, the observables will be described at particle-level. The particle-level is defined after all the partons have hadronized to form stable particles 
with a proper lifetime  above 10\mm/c.

The exclusive jet multiplicity (\Njets) of up to 6 jets (inclusive for $\Njets \geq 7$), in three bins of \ptmax ($200 < \ptmax < 400$, $400 < \ptmax < 800$, and $\ptmax > 800\GeV$) and for three different regions in \dphiOneTwo ($0< \dphiOneTwo < 150$, $150 < \dphiOneTwo < 170$, $170< \dphiOneTwo < 180^\circ$) is measured: 
\begin{equation}
\frac{\rd\sigma_\text{dijet}}{\rd N^\mathrm{i}_\text{jets}\rd p^\mathrm{j}_\mathrm{T1}\rd(\dphiOneTwo^\mathrm{k})}
\end{equation}
where $\mathrm{i},\mathrm{j},\mathrm{k}$ corresponds to the binning in \Njets, \ptmax, and \dphiOneTwo.
We measure exclusive jet multiplicities since we are interested in how the hard and soft multijet radiation is modeled by including higher-order corrections and parton shower effects.   

In addition, the differential cross section as a function of the \pt of each of the four leading jets are measured:
\begin{equation}
\dd{\sigma_{\Pp\Pp\to \mathrm{jj}}}{\ptmax}, \quad \dd{\sigma_{\Pp\Pp\to \mathrm{jj}}}{\ptsec}, \quad \dd{\sigma_{\Pp\Pp\to \mathrm{jjj}}}{\ptthird}, \quad \dd{\sigma_{\Pp\Pp\to \mathrm{jjjj}}}{\ptfourth},
\end{equation}
where $\sigma_{\Pp\Pp\to \mathrm{jj}}$, $\sigma_{\Pp\Pp \to \mathrm{jjj}}$, $\sigma_{\Pp\Pp \to \mathrm{jjjj}}$ correspond to the dijet, three-jet, and four-jet cross sections in proton-proton collision.

\begin{figure*}[h]
	\centering
	{\includegraphics[width=1.0\textwidth]{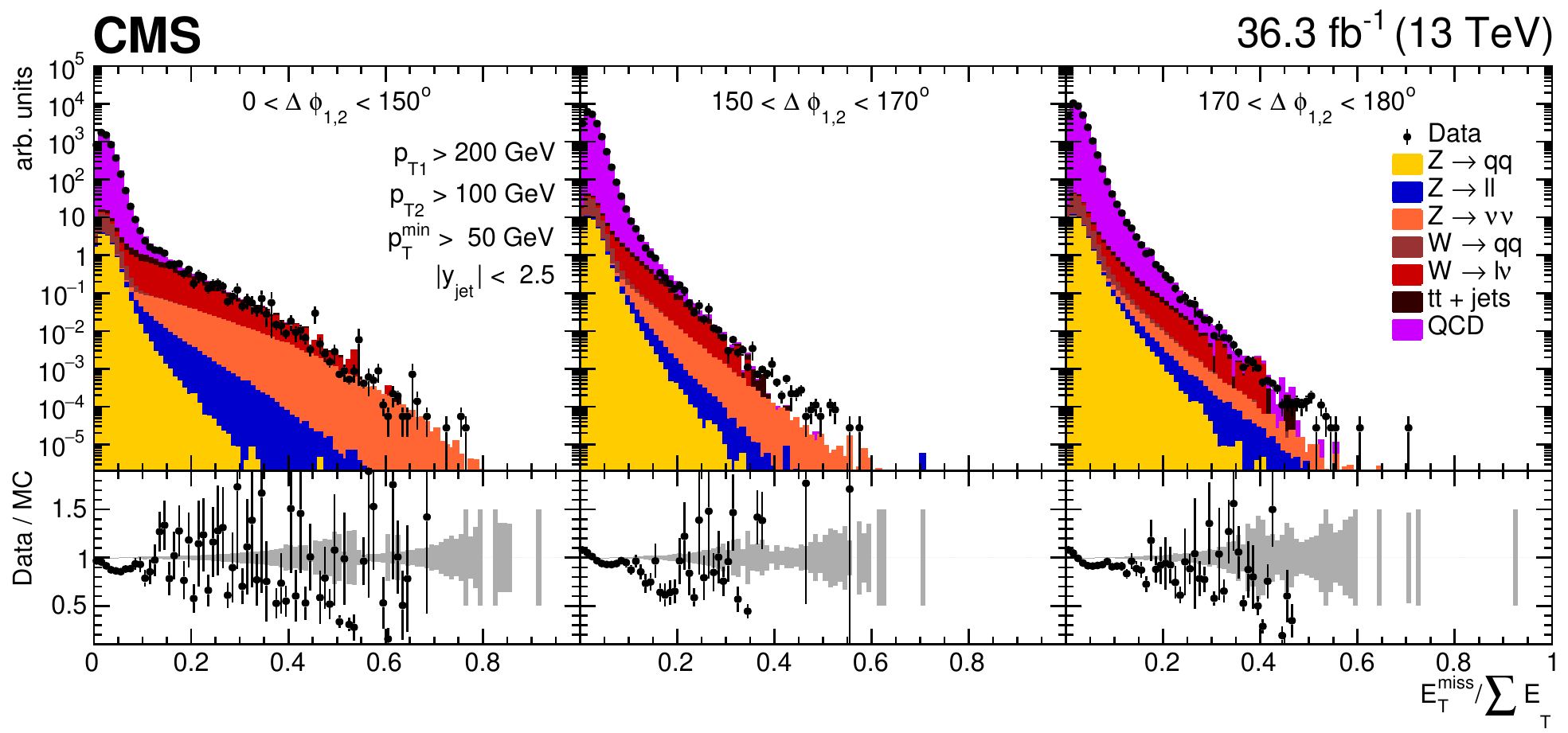}}
	\caption{Distribution of $\ETm/\sum{\ET}$ for data and simulated jet production for three regions of \dphiOneTwo. Shown are the contributions from QCD, $\PW/\PZ$ and $\ttbar$ events. The main contributions of events with large $\ETm$ in the final state come from processes like $\PZ \to  \Pgn \Pagn$ and $\PW \to \Pl\Pgn$. The data (MC prediction) statistical uncertainty is shown as a vertical line (grey shaded bar in the ratio). }  \label{fig:METfraction}
\end{figure*}

\subsection{Background treatment}\label{sec:background}
\begin{tolerant}{900}
To remove background contributions from $\ttbar+\text{jets}$, $\PW/\PZ+\text{jets}$, $\PZ\to \Pgn \Pagn $ and $\PW\to \Pl\Pgn$, events with  $\ETm/\sum{\ET} > 0.1$ are rejected. The missing transverse energy is calculated from \ptmiss\ and the sum $\sum{\ET} $ runs over all PF objects in the event. 
Less than 1\% of the events are rejected in the whole sample. For $\Njets = 2$ this constraint becomes important and the backgournd contribution is reduced from 20\% to the percent level for $\ptmax > 800\GeV$ in the low \dphiOneTwo.
In Fig.~\ref{fig:METfraction}, the  measured fraction $\ETm/\sum{\ET}$ is compared with the simulation of signal and background processes in bins of \dphiOneTwo.
\end{tolerant}

\subsection{Correction for detector effects}\label{sec:correction}
\begin{tolerant}{900}
The measured cross sections are corrected for misidentified events, detector resolution, and inefficiencies for comparison with particle-level predictions through the procedure of unfolding. A response matrix (RM) mapping the ``true'' distribution onto the measured one is constructed using simulated MC samples from \PYTHIAE, \MADGRAPHLO, and \HERWIGpp. The \MADGRAPHLO\ sample, which has the smallest statistical uncertainty, is used as default for constructing the RM, whereas the \HERWIGpp\ and \PYTHIAE\ samples are used to investigate the model dependence.
The RM is constructed by matching the detector and particle-level distributions.
If events (or jets) cannot be matched, they contribute to the background or inefficiency distributions. For the jet multiplicity, the dijet system (leading and subleading jets) is matched if the jets coincide within $\Delta R < 0.2$ (half of the jet radius of  $R = 0.4$). For the \pt\ distributions, jets are matched with $\Delta R < 0.2$, and from the matched candidates the one highest in \pt\ is selected (only events with at least two jets are considered in the matching).
The \textsc{TUnfold} (version 17.9) package~\cite{Schmitt:2012kp} is used to perform the unfolding.
\end{tolerant}

The jet multiplicity is obtained by matrix inversion:
\begin{equation} \label{matrix}
\mathbf{A}\alpha + \beta  =  \gamma
\end{equation}
where $\gamma(\alpha)$ is the distribution at  detector (particle) level, $\mathbf{A}$ is the probability matrix (PM), and $\beta$ is the contribution from background (or noise). The PM is obtained by normalizing the RM to the particle-level axis (row-by-row normalization), and describes the probability that a particle-level jet (or event) generated in a bin is reconstructed in (migrates to) another bin at detector-level.
The condition number of the PM is defined as the ratio between the largest and smallest singular value of the matrix. The condition number of the PM for the jet multiplicity is 3.0, and matrix inversion is possible.

For the  \pt\ distributions, the condition number of the PM is 4.9, and least-square minimisation is applied:
\begin{align} \label{chi2Unf}
\chi^2 &= \min_{\alpha} \left[ (\gamma - \beta - \mathbf{A} \alpha)^\intercal\, {\mathbf{V}_{\gamma\gamma}}^{-1}\, (\gamma - \beta - \mathbf{A}\alpha) \right],
\end{align}
where $\mathbf{V}_{\gamma\gamma}$ represents the statistical covariance matrix of $\gamma$ and $\beta$, with twice the number of
bins at detector level compared to particle-level. No additional regularization is needed.

\begin{figure*}
	\centering
	{\includegraphics[width=1.\textwidth]{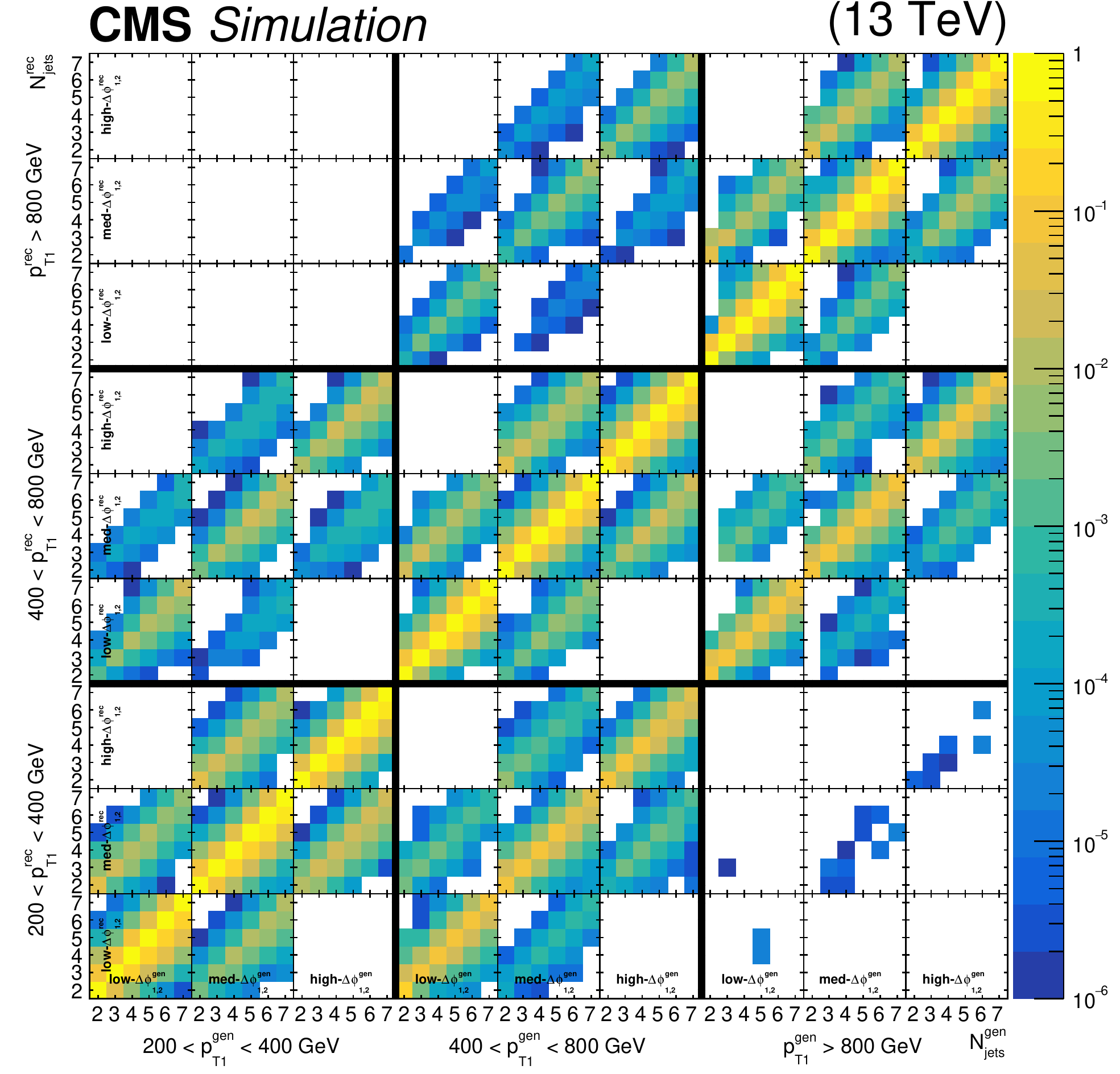}}
   \caption{Probability matrix (condition number: 3.0) for the jet multiplicity distribution constructed with the \MADGRAPHLO sample. The global 3$\times$3 sectors (separated by the thick black lines) correspond to the \ptmax bins, indicated by the labels in the x (lower) and y (left) axes; the  smaller 3$\times$3 structures correspond to the \dphiOneTwo bins, indicated in the leftmost row and lowest column, the x(y) axis of these \dphiOneTwo cells corresponds to the jet multiplicity at particle (detector) level. The z axis covers a range from $10^{-6}$ to $1$ indicating the probability of migrations from the particle-level bin to the corresponding detector-level bin. The lower right corner of the matrix, which describes (extreme) migrations between hight \ptmax at particle level and low \ptmax at dectector level, is characterized by statistical fluctuations.}
    \label{fig:NjPhiptmax_PM}
\end{figure*}

\begin{figure*}
	\centering
	{\includegraphics[width=1.\textwidth]{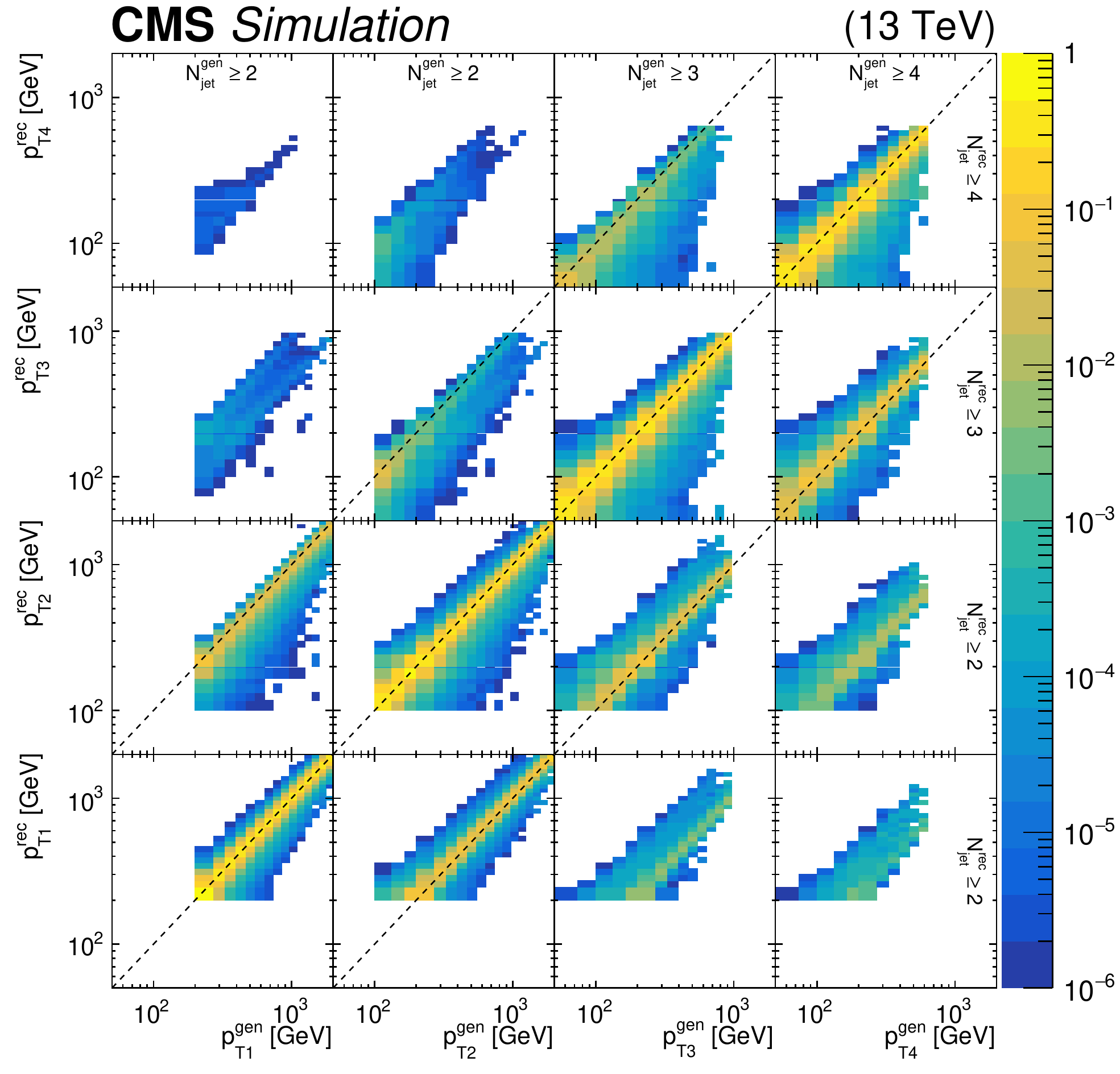}}
	\caption{Probability matrix (condition number: 4.9) for the \pt of the four leading jets constructed with the \MADGRAPHLO sample. The global 4$\times$4 sectors correspond to the \pt distributions each of the first four jets, the x axis corresponds to the particle (gen) level and y axis corresponds to the detector (rec) level. The z axis covers a range from $10^{-6}$ to $1$ indicating the probability of migrations from the particle-level bin to the corresponding detector-level bin.}
	\label{fig:PtN_PM}
\end{figure*}

In Fig.~\ref{fig:NjPhiptmax_PM} and~\ref{fig:PtN_PM}, the PMs are shown for the multiplicity distributions and the \pt\ distributions of the first four jets.

\subsection{Uncertainties}\label{sec:uncertainties}
\begin{figure*}
	\centering	
	{\includegraphics[width=1.\textwidth]{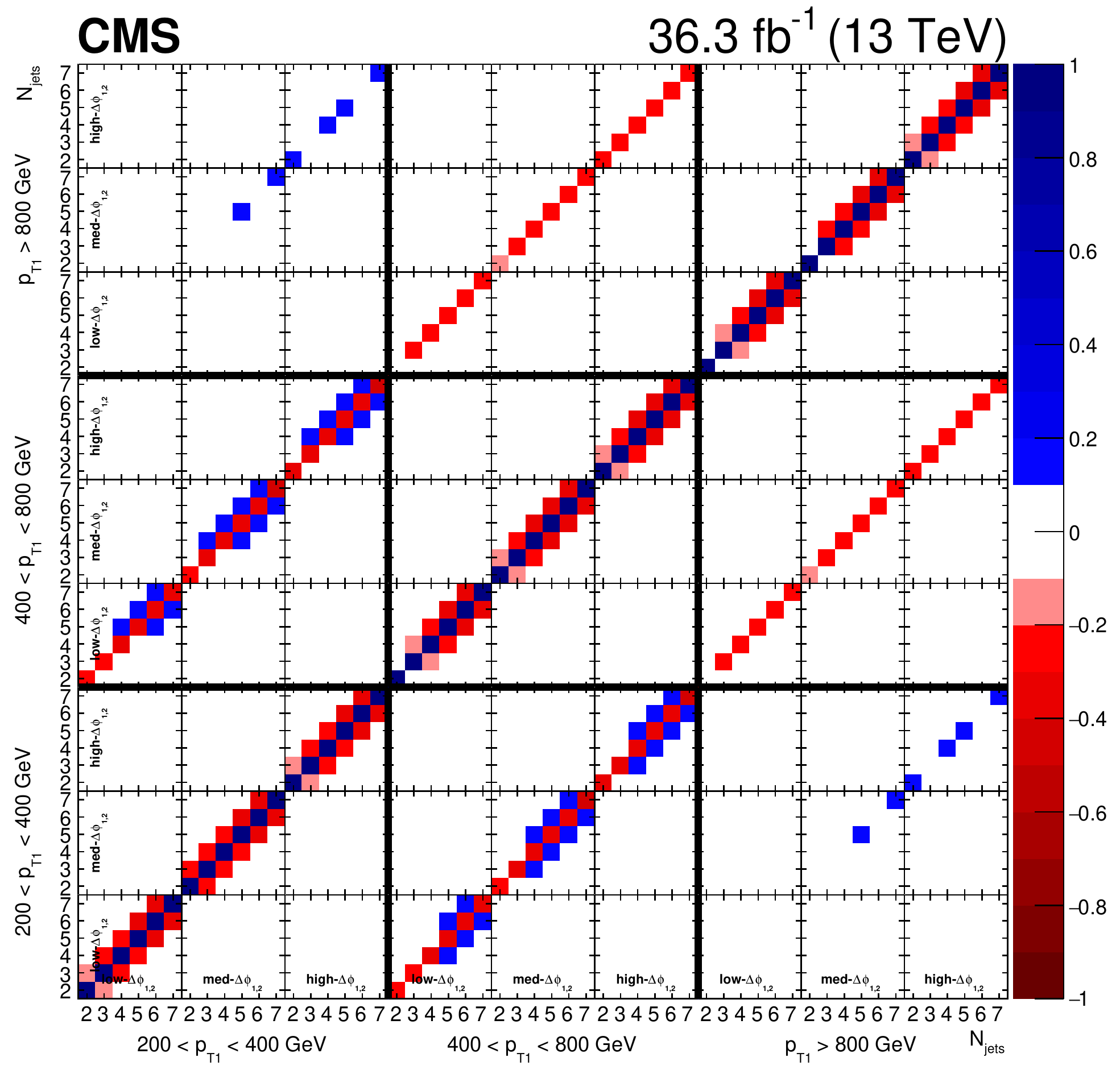}}
	\caption{Correlation matrix at the particle-level for the jet multiplicity distribution. It contains contributions from the data and from the limited-size \MADGRAPHLO sample. The global 3$\times$3 sectors (separated by the thick black lines) correspond to the \ptmax bins, indicated by the labels next to the x (lower) and y (left) axes; the smaller 3$\times$3 structures correspond to the \dphiOneTwo bins, indicated in the leftmost row and lowest column, the x and y axes correspond to the jet multiplicity. The z axis covers a range from $-1$ to $1$ indicating the correlations in blue shades and anticorrelations in red shades, the values between $-0.1$ and $0.1$ are represented in white.}
	\label{fig:NjPhiptmax_CorrM}
\end{figure*}

\begin{figure*}
	\centering
	{\includegraphics[width=1.\textwidth]{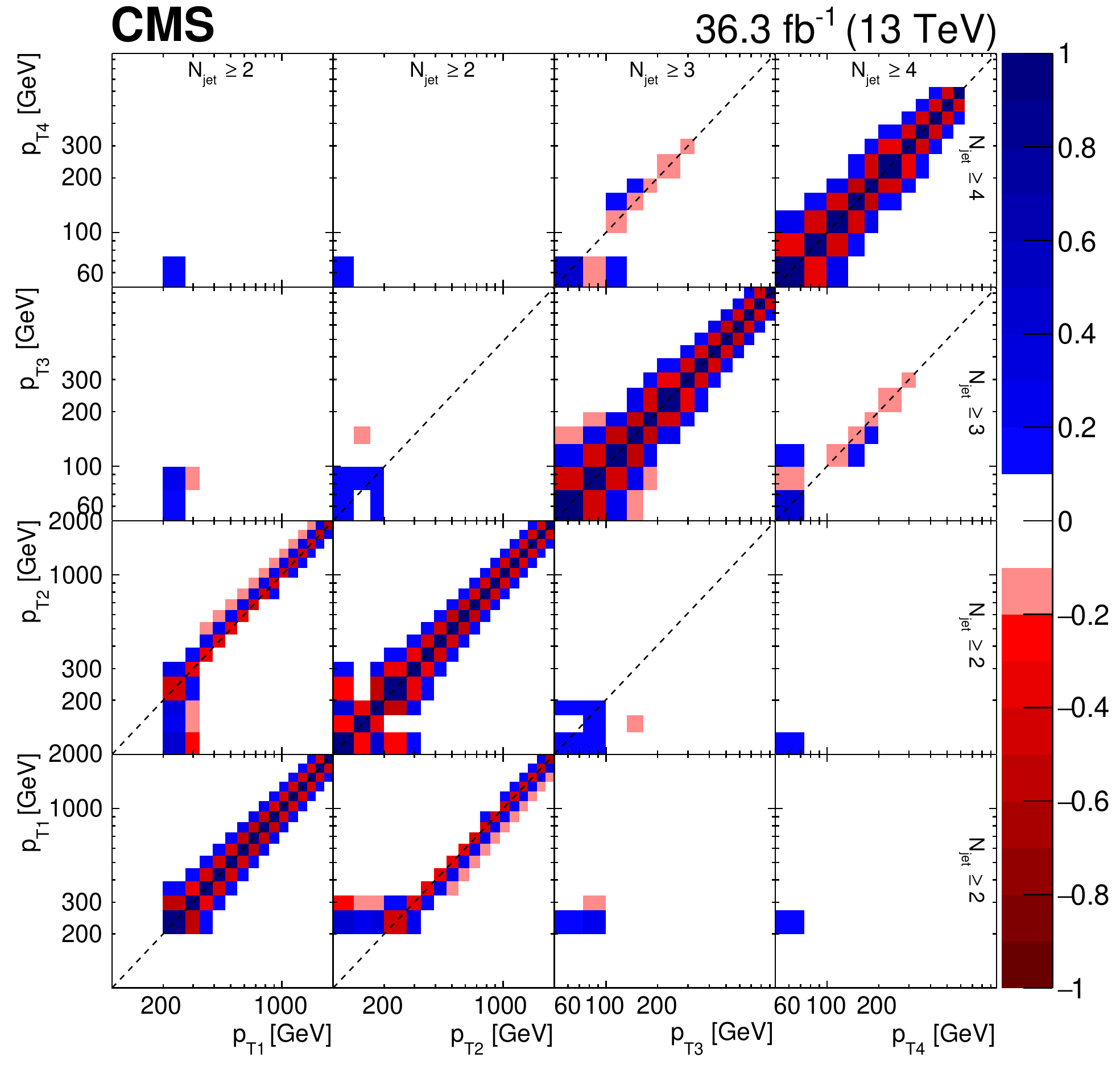}}
	\caption{Correlation matrix for the particle-level \pt of the four leading jets. It contains contributions from the data and from the limited-size  \MADGRAPHLO sample. Here each one of the 4$\times$4 sectors corresponds to one of the \pt spectra measured, indicated by the x and y axis labels. The z axis covers a range from $-1$ to $1$ indicating the correlations in blue shades and anticorrelations in red shades, the values between $-0.1$ and $0.1$ are represented in white.}
	\label{fig:PtN_CorrM}
\end{figure*}

The statistical uncertainties in the measured spectra and response matrices are propagated to the final results. 
In Fig.~\ref{fig:NjPhiptmax_CorrM} and~\ref{fig:PtN_CorrM}, the statistical correlations are shown for the jet multiplicity and for the \pt spectra of the four leading jets. 

The systematic uncertainties originate from the following sources:
\begin{itemize}
\item JES: 
The JES uncertainty is estimated from variations of one standard deviation in the JES corrections applied to data (at detector-level) and repeating the whole unfolding procedure for each variation.
\item JER: 
The JER uncertainty is estimated by varying the resolution by one standard deviation in the simulated sample, and repeating the unfolding for each variation.
\item Integrated luminosity: 
The uncertainty in the integrated luminosity is 1.2\%~\cite{CMS:2021xjt} and is applied as a global scaling factor to the cross section.
\item Pileup: 
The pileup distribution in the simulated samples is reweighted to match that of the data. The corresponding uncertainty is estimated by varying the total inelastic cross section by $\pm5\%$~\cite{CMS:2018mlc}, affecting the measurement by less than 1\%.
\item Trigger prefiring uncertainty: 
The trigger prefiring uncertainty is estimated by varying the correction applied to the simulated samples 
and repeating the unfolding for each variation, resulting in an uncertainty of 1 to 3\%.
\item Model uncertainty: 
The model uncertainty is estimated by varying the distributions of the factorization and renormalization scales in the MC sample such as to still describe the detector level distributions. Additionally background and inefficiencies are varied separately by 15\%. The final uncertainty is the quadratic sum of each of the uncertainties.
It was validated with a set of closure tests performed using \PYTHIAE and \HERWIGpp samples as pseudodata unfolded with \MADGRAPHLO RM. This uncertainty ranges from 1 to 7\%.
\end{itemize}
The total systematic uncertainty is obtained by adding all the systematic uncertainties in quadrature, assuming independent sources.

\begin{figure*}
	\centering
	{\includegraphics[width=1.\textwidth]{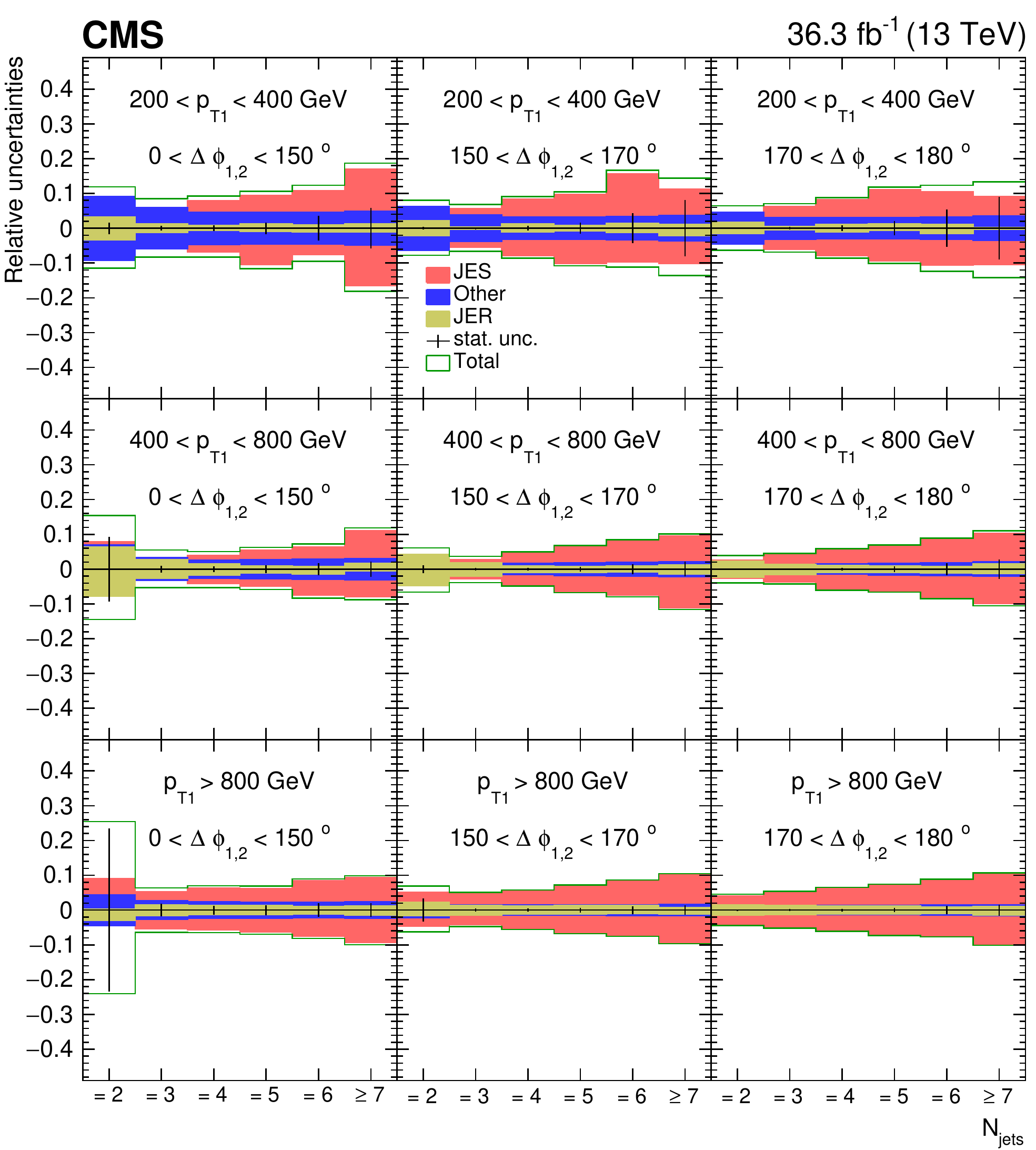}}
	\caption{Relative uncertainties for JES, JER, ``Other'' and total statistical uncertainty  for the jet multiplicity distribution in bins of \ptmax and \dphiOneTwo. Here ``Other'' indicates luminosity, pileup, prefiring, and model uncertainty.}
	\label{fig:MultUnc}
\end{figure*}

In Fig.~\ref{fig:MultUnc}, the relative uncertainties for the jet multiplicity in bins of \ptmax and \dphiOneTwo\ are shown. The dominant uncertainty is JES. The total statistical uncertainty ({stat. unc.}) is mainly driven by the limited event counts in data. 
The total experimental uncertainty ({Total}) is typically about $10$ to $15$\%.

\begin{figure*}
	\centering
	{\includegraphics[width=1.\textwidth]{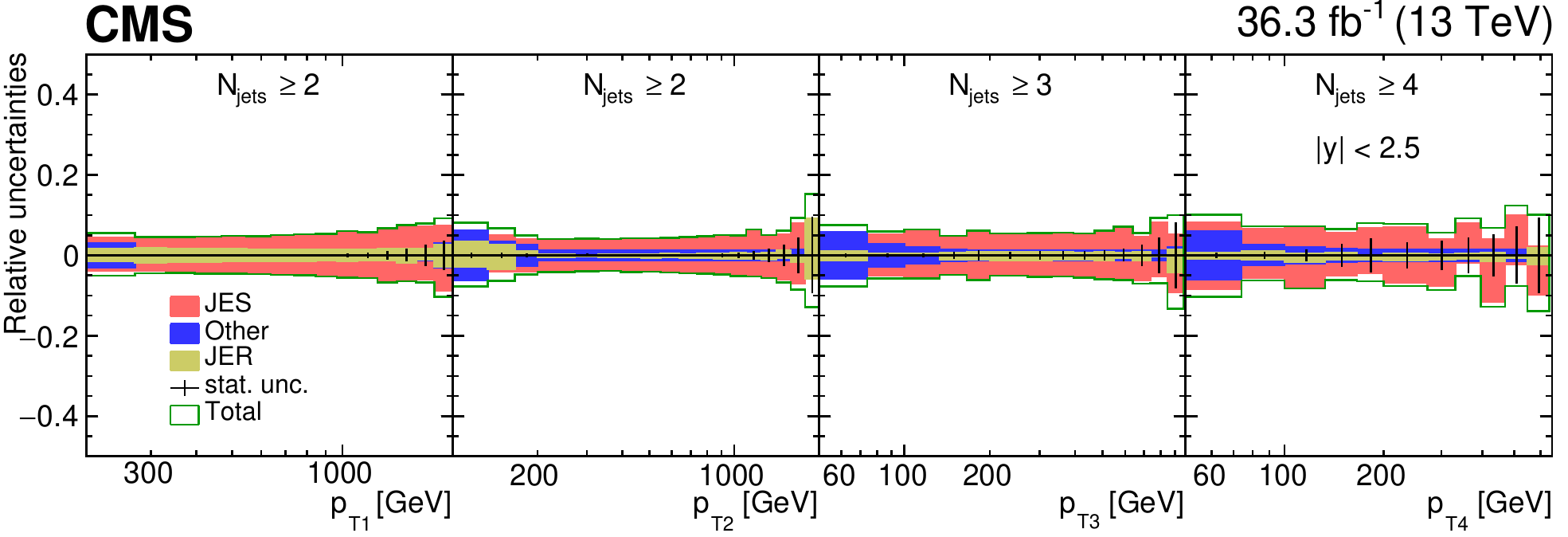}}
	\caption{Relative uncertainties for JES, JER, ``Other'' and total statistical uncertainty for the \pt distributions of the four leading jets. Here ``Other'' indicates luminosity, pileup, prefiring, and model uncertainty.}
	\label{fig:PtNUnc}
\end{figure*}

In Fig.~\ref{fig:PtNUnc}, the relative uncertainties as a function of the jet \pt  for the four leading jets are shown. The dominant uncertainty is {JES}. The measurement is limited by the systematic uncertainty and the total experimental uncertainty ranges from $ 5$ to $10$\%.
 
\section{Results} \label{sect:results}

The phase space of the measurements at particle-level (particles with a proper lifetime above 10\mm/c) 
is defined by jets clustered using the anti-\kt algorithm~\cite{Cacciari:2008gp, Cacciari:2011ma} with a distance parameter of $R=0.4$ within  
 $\abs{y} < 3.2$. Events are selected  if the two highest \pt\ jets with $\ptmax > 200\GeV$, $\ptsec > 100\GeV$ have $\abs{y}< 2.5$
 (\ie, events are not counted where one of the leading jets has $\abs{y}> 2.5$).
 For the additional jets  $\pt >50\GeV$ and $\abs{y}< 2.5$ is required.
All predictions (LO and NLO) are normalized to the measured dijet cross section in the measurement phase space, with the normalization factors explicitly given in the figures.

\subsection{Jet multiplicity distribution}
\label{subsec:resultsMultiplicity}
\begin{figure*}    
	
	\centering
	{\includegraphics[width=0.9\textwidth]{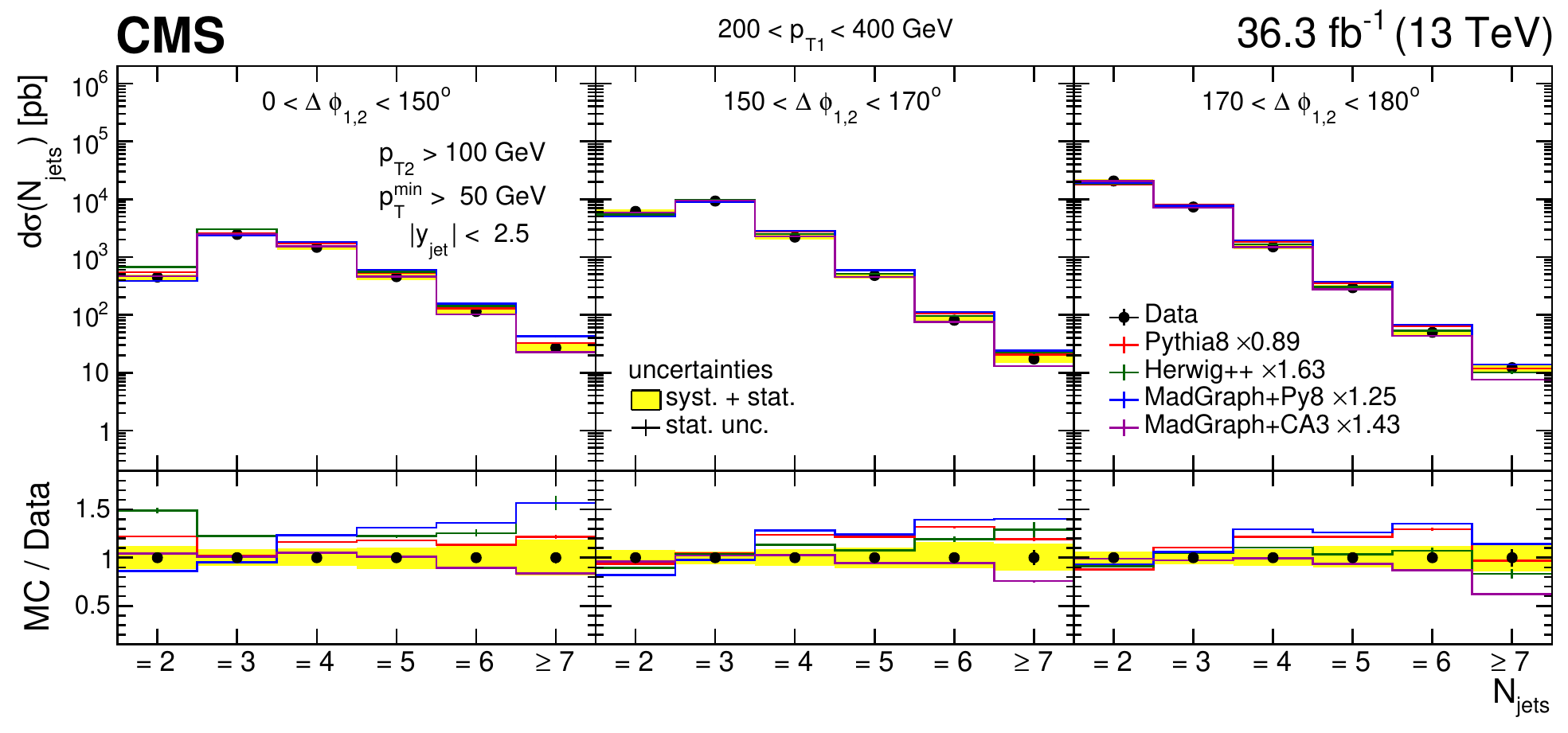}}	
	{\includegraphics[width=0.9\textwidth]{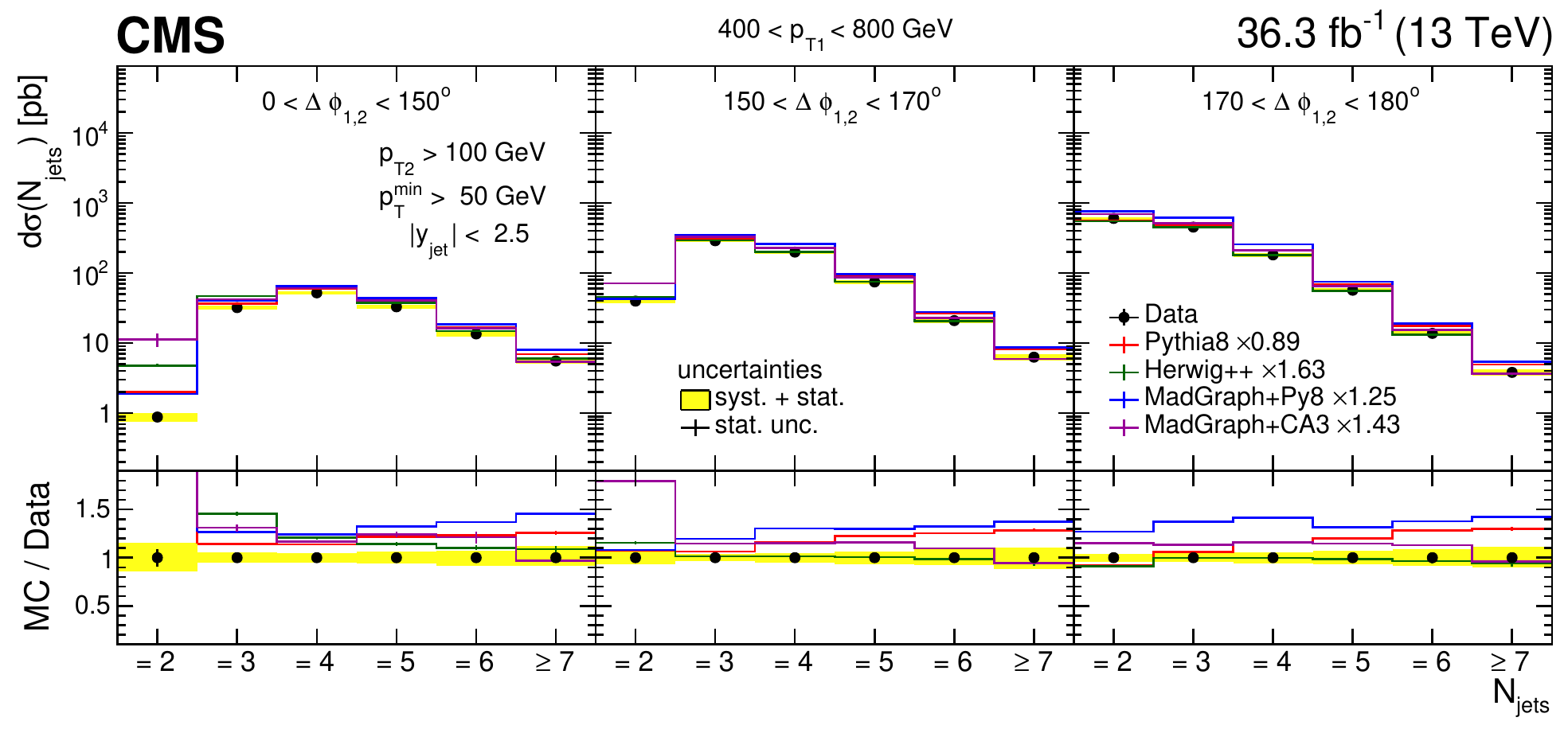}}	
	{\includegraphics[width=0.9\textwidth]{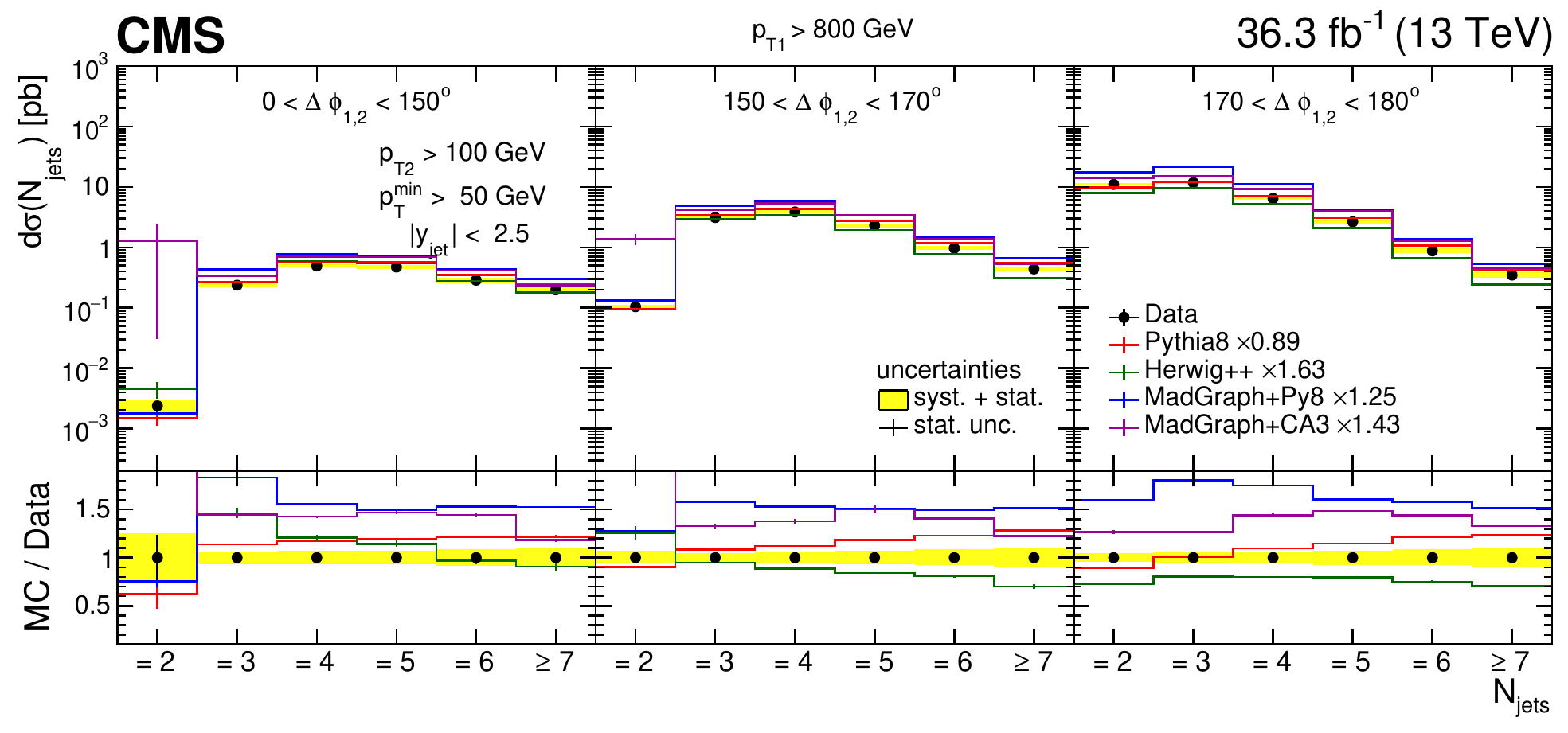}}
	\caption{Differential cross section as a function of the exclusive jet multiplicity (inclusive for 7 jets) in bins of \ptmax and \dphiOneTwo. The data are compared with LO predictions of  \PYTHIAE, \HERWIGpp, \MADGRAPHLO and \MADGRAPHCAS. The predictions are normalized to the measured dijet cross section using the scaling factors shown in the legend. 
	The vertical error bars correspond to the statistical uncertainty, the yellow band shows the total experimental uncertainty. }\label{fig:resultsMultiplicity} 
\end{figure*}

The multiplicity of jets with  $\pt > 50\GeV$ in $\abs{y}< 2.5 $ is measured for various regions of the transverse momentum of the leading jet, \ptmax, and the azimuthal angle \dphiOneTwo\ between the two leading jets as shown in Fig.~\ref{fig:resultsMultiplicity}.

As a characterization of the jet multiplicity we compare the  production rate for 3 jets with the one for 7 jets. 
In the region of low \ptmax ($200 < \ptmax < 400\GeV$), a large number of additional jets is observed at low \dphiOneTwo ($ 0 < \dphiOneTwo < 150^\circ$), the production rate between 3 and 7 jets changes by two orders of magnitude. In the large-\dphiOneTwo region ($ 170 <  \dphiOneTwo < 180^\circ$), where the leading jets are nearly back-to-back, the production rate for 3 to 7 jets changes by three orders of magnitude. We note that even in this back-to-back region a large number of additional jets are observed that do not contribute significantly to the momentum imbalance of the two leading jets.

In the region of large \ptmax ($ \ptmax >800\GeV$), we observe that the rate of additional jets at low \dphiOneTwo is essentially constant and the rate between 3 and 7 jets only shows small changes, indicating that many jets participate in the compensation of the \dphiOneTwo decorrelation. In the large-\dphiOneTwo region ($ 170 <  \dphiOneTwo < 180^\circ$), the rate between 3 and 7 jets changes by less than 2 orders of magnitude, in contrast to the low-\ptmax region. 
A multiplicity  of more than two or three additional jets at large \ptmax is observed in all regions of \dphiOneTwo . 

In Fig.~\ref{fig:resultsMultiplicity}, predictions from the LO $2\to2$ generators \PYTHIAE and \HERWIGpp including parton showering and MPI are shown. The shape of the prediction coming from \PYTHIAE is different from what is observed in the measurement. The shape of the prediction from  \HERWIGpp agrees rather well with the measurement, especially in  the large \dphiOneTwo region. The difference between \PYTHIAE and \HERWIGpp in jet multiplicity comes from the different treatment of the parton shower.
In addition the predictions from \MADGRAPHLO and \MADGRAPHCAS with additional noncollinear high-\pt partons, supplemented with different parton showering approaches and MPI are shown. We find that the merged prediction from \MADGRAPHLO does not agree in shape with the measurement, whereas the \MADGRAPHCAS merged prediction (for $\Njets > 2$ and $\ptmax < 800 \GeV$) does, similarly to the case of \HERWIGpp. 

\begin{figure*}    
	\centering
	{\includegraphics[width=0.9\textwidth]{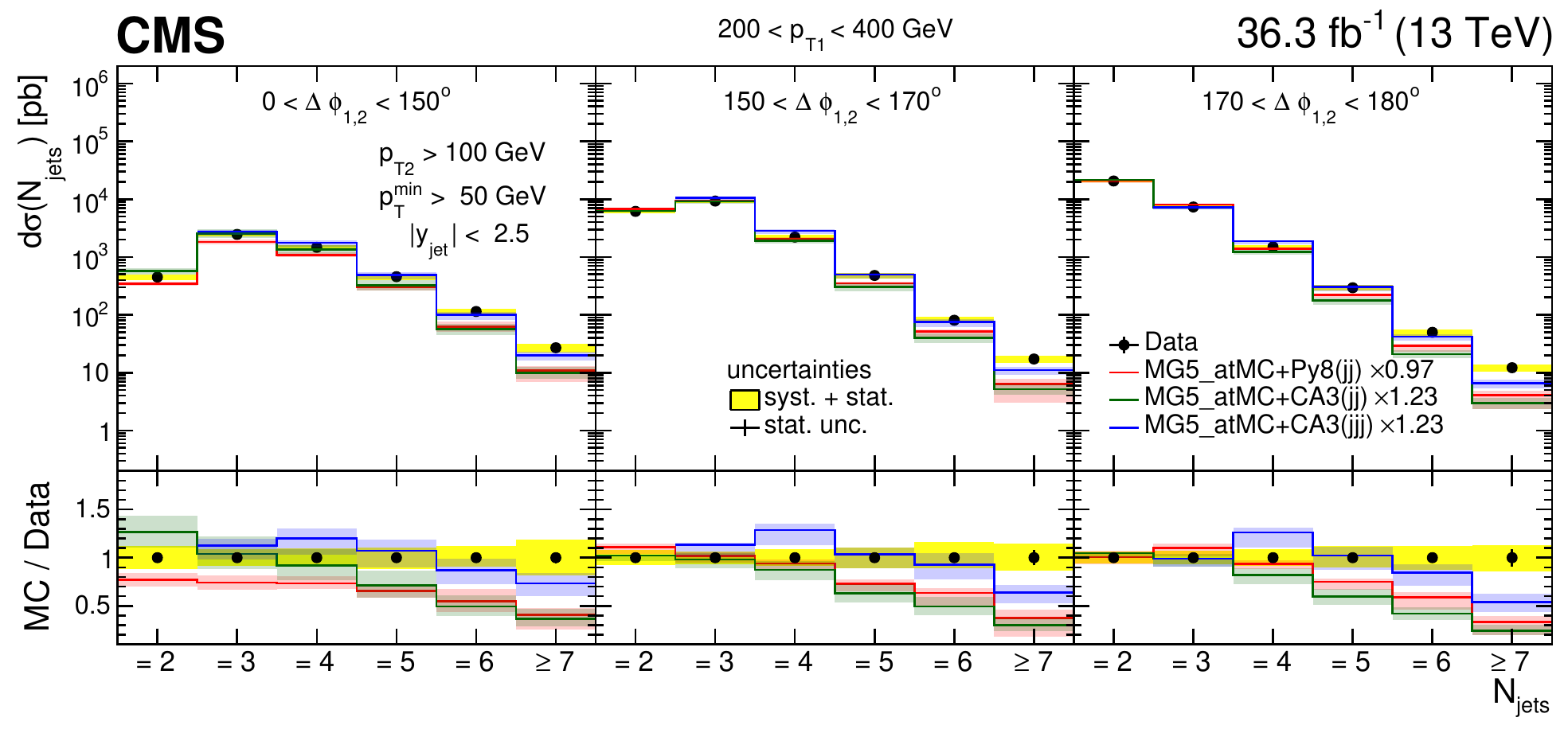}}	
	{\includegraphics[width=0.9\textwidth]{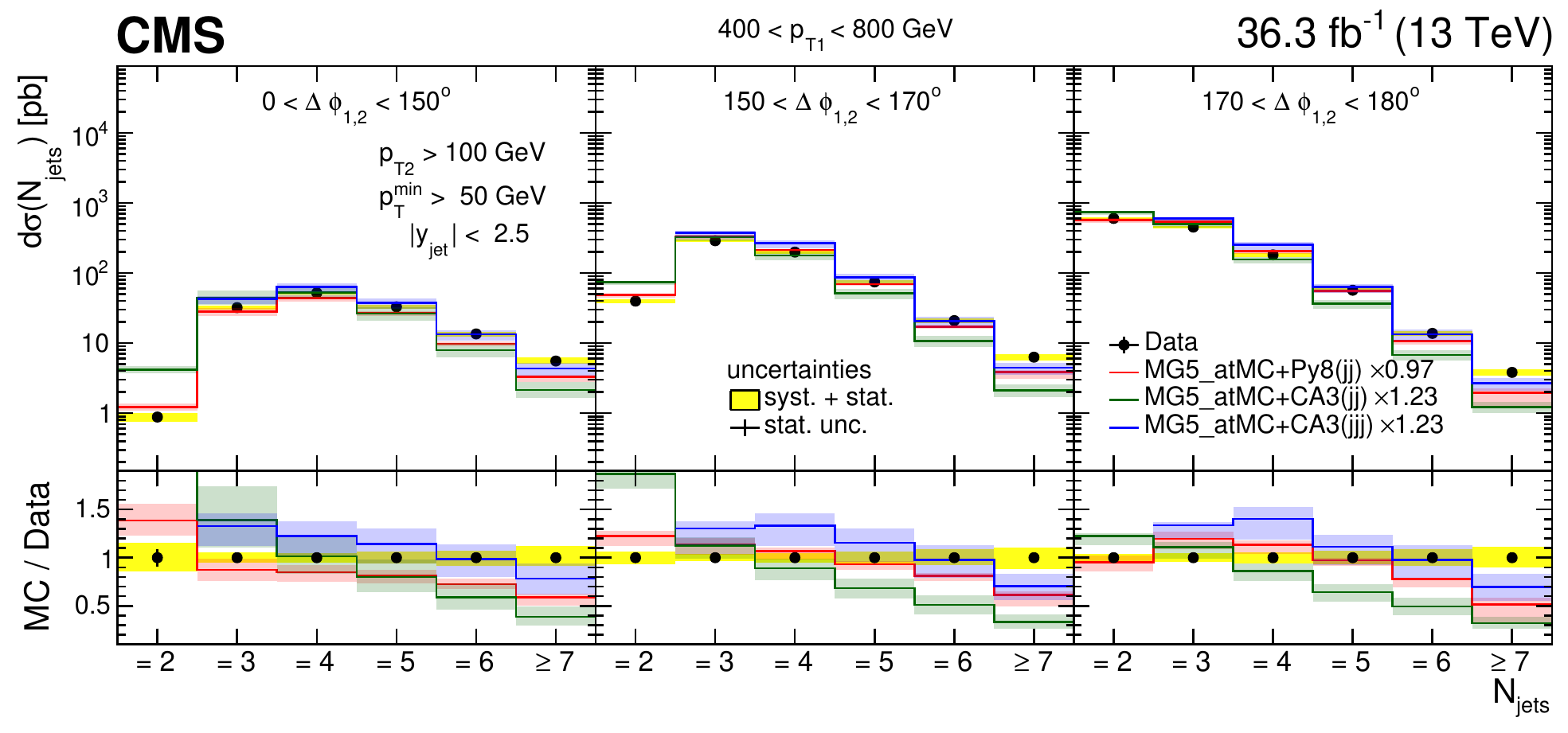}}
	{\includegraphics[width=0.9\textwidth]{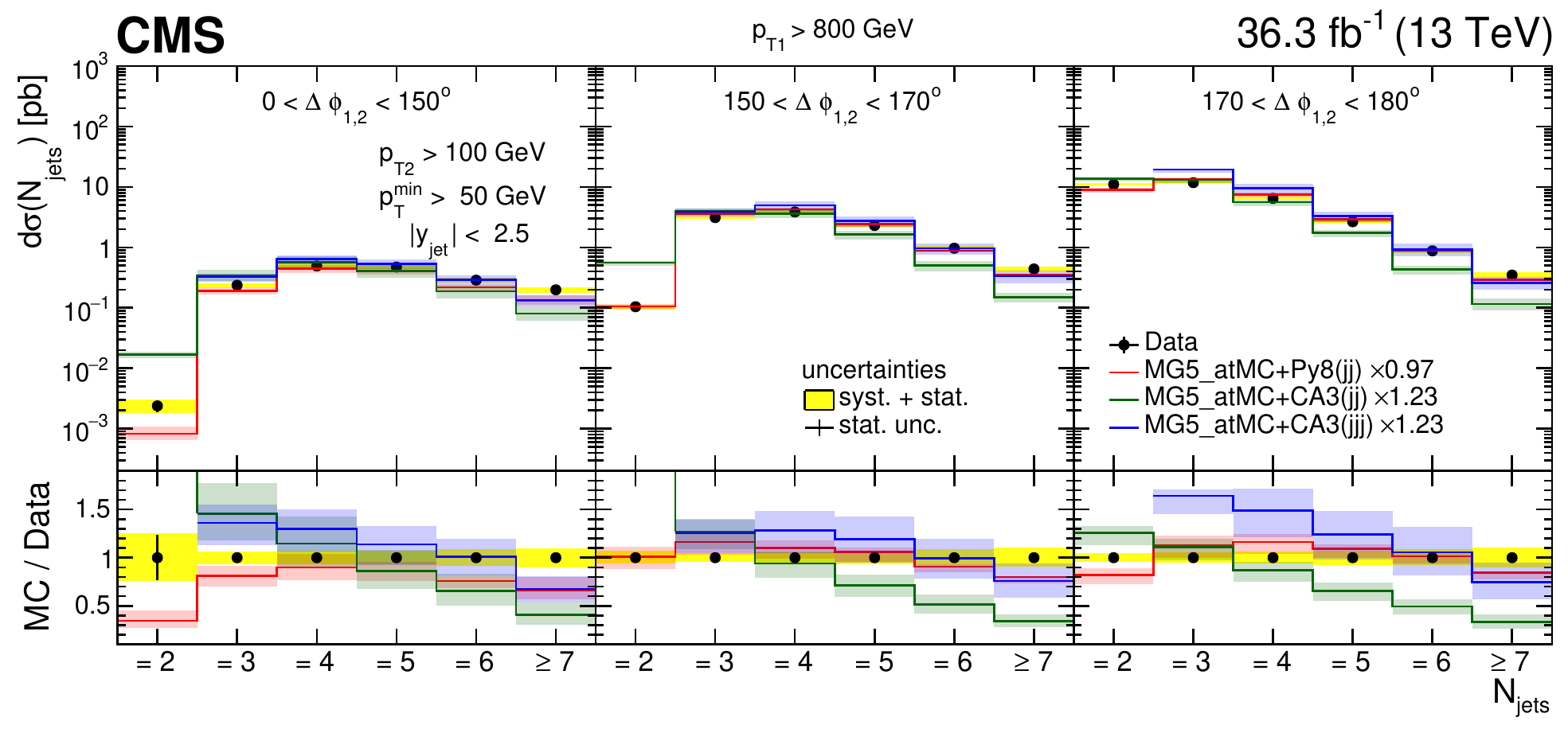}}
	\caption{Differential cross section as a function of the exclusive jet multiplicity (inclusive for 7 jets) in bins of \ptmax and \dphiOneTwo. The data are compared with NLO dijet predictions \aMCPy (jj) and \aMCcas (jj) as well as the NLO three-jet prediction of \aMCcas (jjj). 
	The vertical error bars correspond to the statistical uncertainty, the yellow band shows the total experimental uncertainty. 
	The shaded bands show the uncertainty from a variation of the renormalization and factorization scales. The predictions are normalized to the measured inclusive dijet cross section using the scaling factors shown in the legend.}\label{fig:resultsMultiplicityNLO} 
\end{figure*}

The calculations with NLO MEs matched with parton shower compared with the measurements are shown in Fig.~\ref{fig:resultsMultiplicityNLO}. The uncertainty bands of the predictions come from the variation of the factorization and renormalization scales by a factor of two up and down, avoiding the largest scale differences.
The normalization of the \aMCPy (jj) calculation is in reasonable agreement with the measured cross section even for three jets.
For higher jet multiplicities the prediction falls below the measurement.
\aMCcas (jj) predicts a smaller cross section for more than three jets compared with the measurement. The \aMCcas (jjj) NLO calculation (using the same normalization factor as for \aMCcas (jj)) predicts a larger three- and four-jet cross section, whereas the higher jet multiplicities are still underestimated. 

\subsection{Transverse momenta of the four leading jets}
\label{subsec:resultsPt}

\begin{figure*}    
	\centering
	{\includegraphics[width=1.\textwidth]{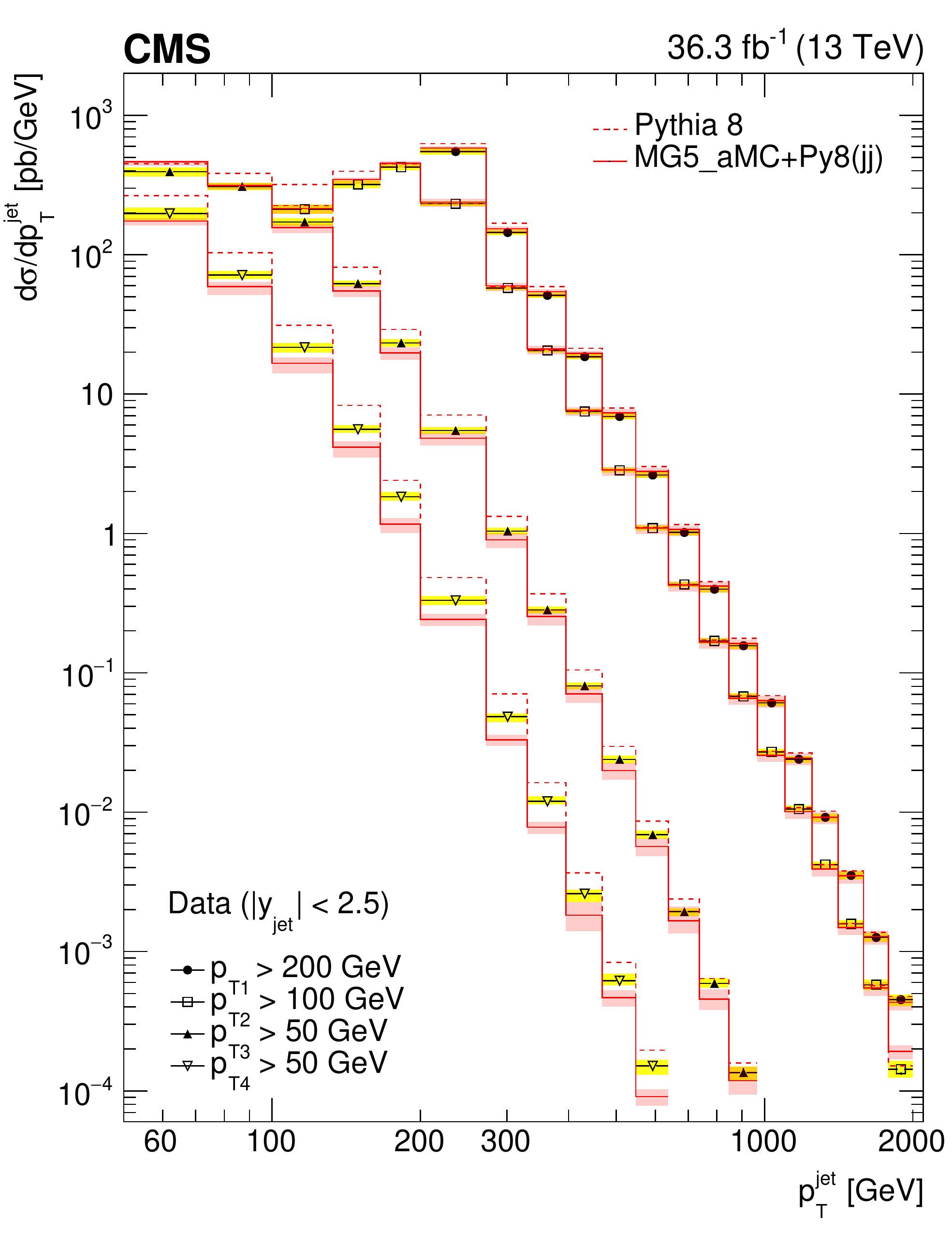}} 	
	\caption{Transverse momenta of the  four leading jets, with the yellow band representing the total experimental uncertainty. The data are compared with LO (\PYTHIAE) and NLO (\aMCPy) predictions. The red band in the NLO prediction represents the renormalization and factorization scale uncertainty.
	}\label{fig:resultsPtNData} 
\end{figure*}

The  measured differential jet cross section as a function of the jet transverse momentum, \pt, for the four leading \pt\  jets is shown in Fig.~\ref{fig:resultsPtNData} and compared with the predictions of  \PYTHIAE and \aMCPy (jj).   
The \pt values of the two leading-\pt jets reaches up to 2\TeV and the third and fourth jets \pt reaches about 1\TeV. We observe that the shape of the \pt spectrum  for the third and fourth leading jets is qualitatively similar to the one of the two leading jets, whereas the cross section is different. The 
rise of the cross section for the second jet
between 100\GeV and 200\GeV is a consequence of the higher \pt requirement ($\ptmax > 200\GeV$) applied to the leading jet in the event selection.

\begin{figure*}    
	\centering
	{\includegraphics[width=0.9\textwidth]{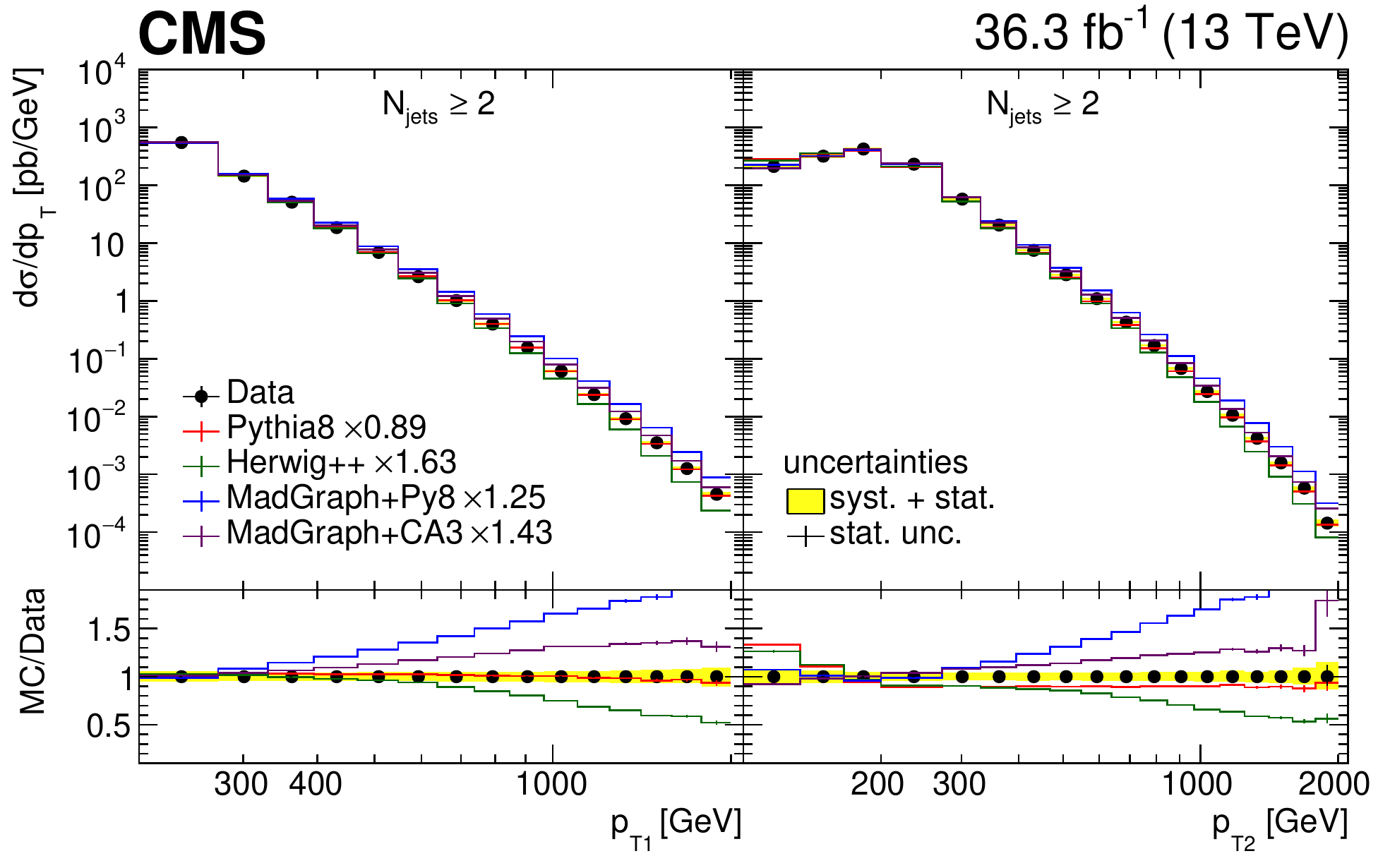}}	
	{\includegraphics[width=0.9\textwidth]{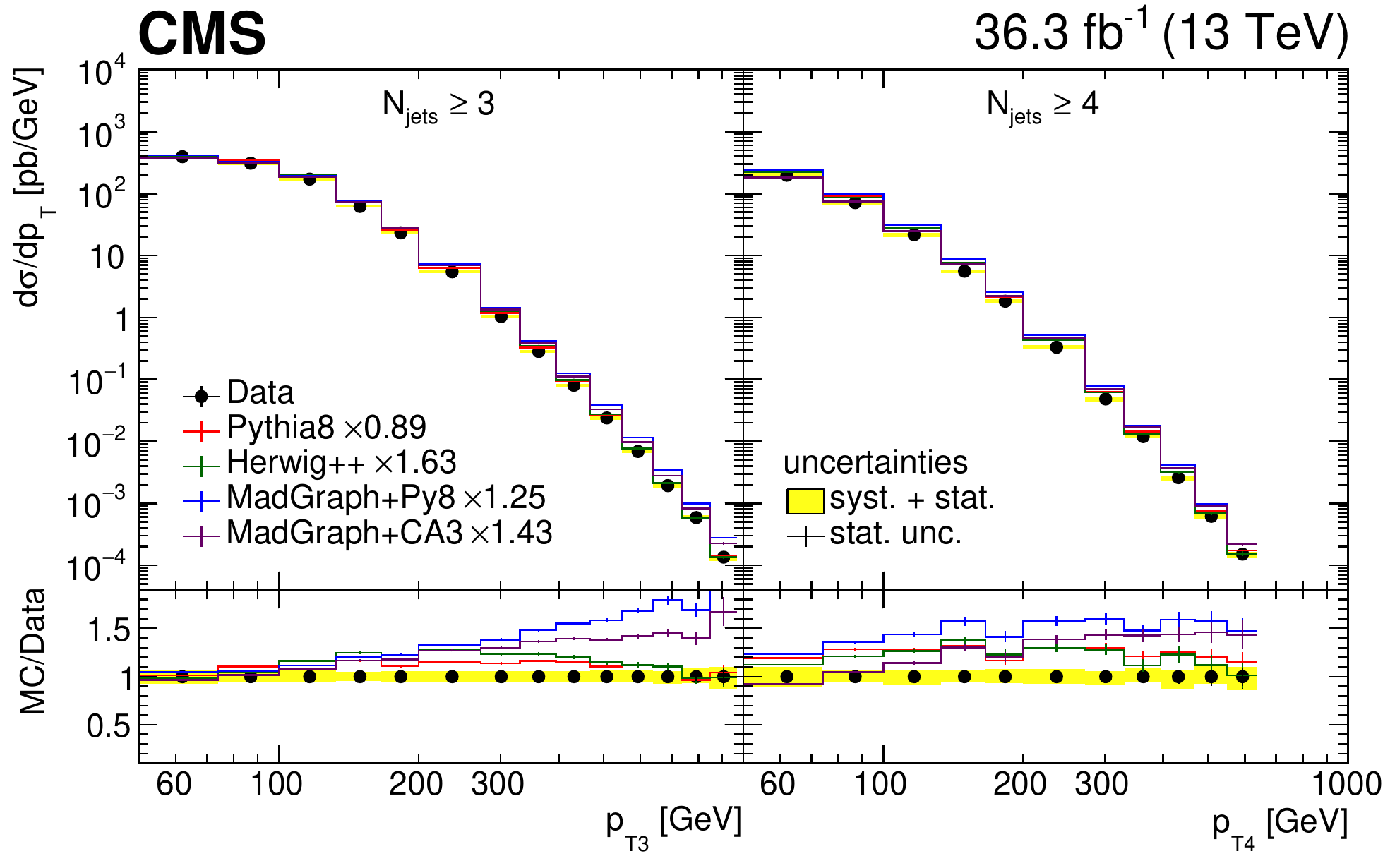}}
	\caption{Transverse momentum distributions of the four leading jets. The transverse momentum of the leading and subleading (third and fourth leading) \pt jets from left to right is shown in the upper (lower) figure. The data are compared with LO predictions of \PYTHIAE, \HERWIGpp, \MADGRAPHLO and \MADGRAPHCAS. The predictions are normalized to the measured dijet cross section using the scaling factors shown in the legend. 
	The vertical error bars correspond to the statistical uncertainty, the yellow band shows the total experimental uncertainty. 
	}\label{fig:resultsPt} 
\end{figure*}

In Fig.~\ref{fig:resultsPt}, the measured differential cross section as a function of the \pt for the four leading jets is shown and compared with LO predictions (using the same normalization factors as in Fig.~\ref{fig:resultsMultiplicity}). 
Only the prediction of \PYTHIAE is able to describe reasonably well the measurement in shape, except for the region $\ptsec < 200\GeV$. Among the LO calculations, \PYTHIAE and \HERWIGpp predict a rather flat distribution for the third and fourth jet, whereas the other predictions are different in shape.
The predictions from \HERWIGpp are not in agreement in shape and rate with the measurements, the differences are up to 50\% for the leading and subleading jets at large \pt .
The prediction from \MADGRAPHLO gives a significantly different shape for the \pt spectrum for the first 3 jets. Comparing the merged predicitons, \MADGRAPHCAS gives a significant improvement in the description of the shape of the \pt distribution for the three leading jets, whereas the description of the distribution of the fourth jet is similar to the one with \MADGRAPHLO .

\begin{figure*}    
	\centering
	{\includegraphics[width=0.9\textwidth]{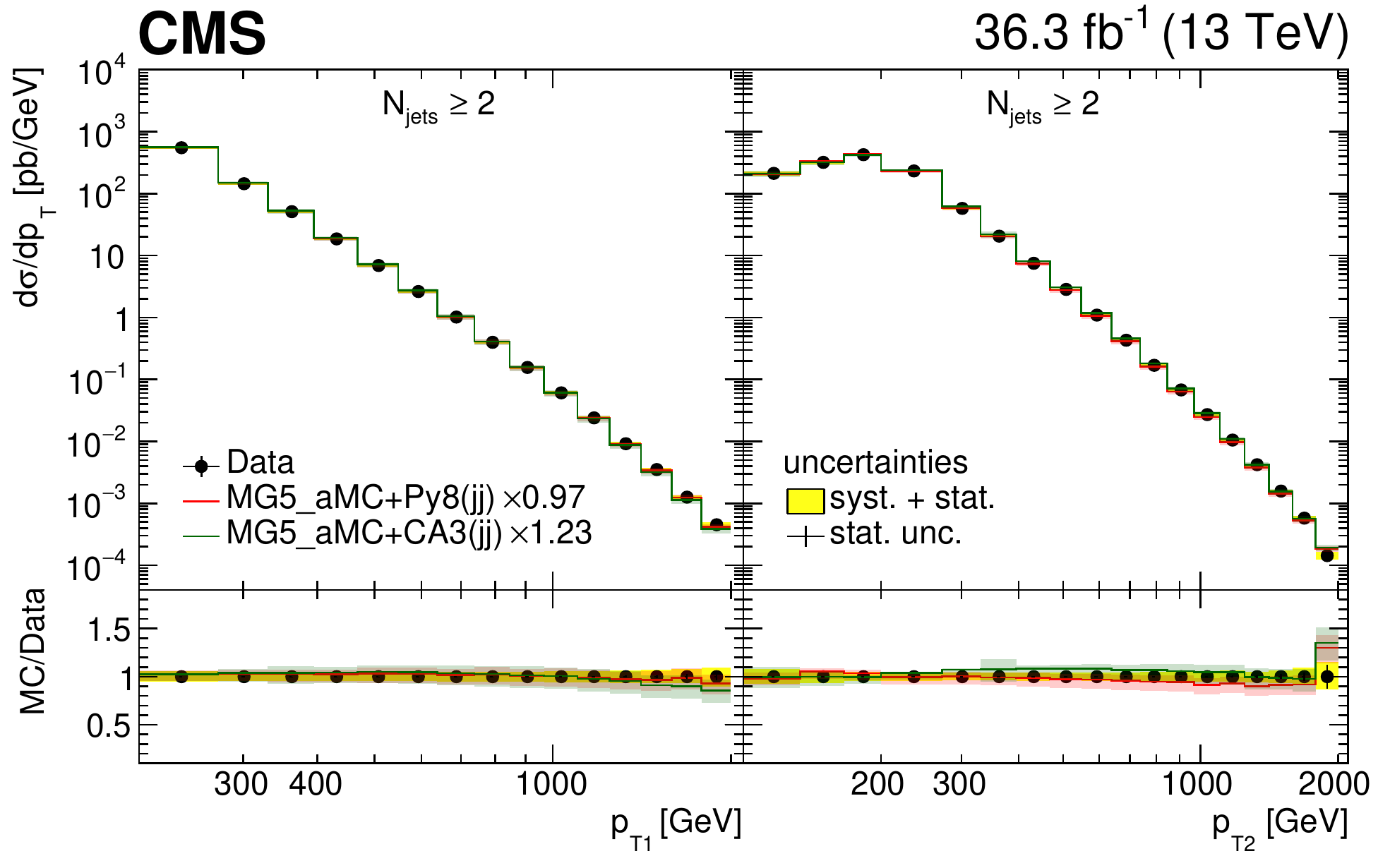}}	
	{\includegraphics[width=0.9\textwidth]{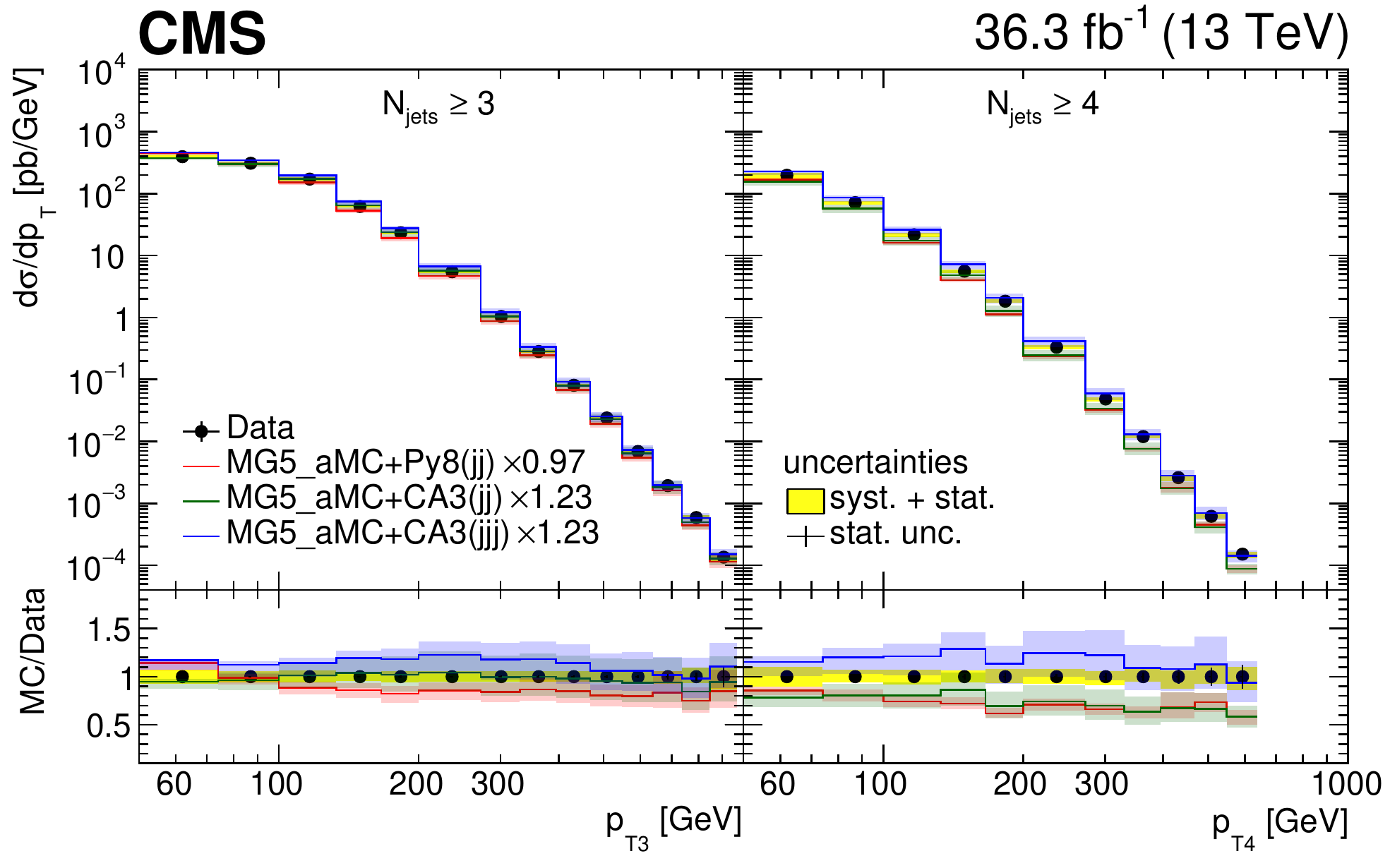}}
	\caption{Transverse momentum distributions of the four leading jets. The transverse momentum of the leading and subleading (third and fourth leading) \pt jets from left to right is shown in the upper (lower) figure. The data are compared with NLO predictions \aMCPy (jj) and \aMCcas (jj) as well as the NLO three-jet prediction of \aMCcas (jjj). 
	The vertical error bars correspond to the statistical uncertainty, and the yellow band to total uncertainty of the measurement. 
	The bands show the uncertainty from a variation of the renormalization and factorization scales. The predictions are normalized to the measured dijet cross section using the scaling factors shown in the legend. 
}\label{fig:resultsPtNLO} 
\end{figure*}

\begin{tolerant}{900}
The predictions obtained with NLO MEs are shown in Fig.~\ref{fig:resultsPtNLO} using the same normalization factors as in Fig.~\ref{fig:resultsMultiplicityNLO}. The uncertainty bands of the predictions come from the variation of the factorization and renormalization scales by a factor of two up and down, avoiding the largest scale differences.
The predictions of \aMCPy (jj) and \aMCcas (jj) describe the normalization and the shape of the first three jets rather well, whereas for the fourth jet (which comes from the parton shower) falls below the measurement by 20--30\%.
The prediction of \aMCcas (jjj) describes the third and fourth jets rather well within uncertainties (predictions for the first and second jet are meaningless for \aMCcas (jjj) and therefore not shown).
The calculations using \PBM -TMDs together with NLO MEs in the MC@NLO frame provide a good description of jet measurements over a wide range in transverse momentum and jet multiplicities without tunable parameters in the initial-state parton shower. 
\end{tolerant}

\section{Summary}
A study of multijet events has been performed in proton-proton collisions at a center-of-mass energy of 13\TeV using data collected with the CMS detector corresponding to an integrated luminosity of $36.3\fbinv$. The measurements are performed by selecting a dijet system containing a jet with $\pt > 200\GeV$ and a subleading jet with $\pt>100\GeV$ within $\abs{y} < 2.5$.

For the first time, the jet multiplicity in bins of the leading jet \pt and the azimuthal angle difference between the two leading jets, \dphiOneTwo, is measured.  
The jet multiplicity distributions show that even in the back-to-back region of the dijet system, up to seven jets are measurable. The differential production cross sections are measured for the highest \pt jets  up to the \TeV scale.  

The measurement of the differential cross section as a function of the jet \pt for the four highest \pt jets is an important reference for QCD multijet cross section calculations, and especially for the simulations including parton showers for higher jet multiplicity.

The measured multiplicity distribution of jets with $\pt > 50\GeV$ and $\abs{y} < 2.5$ is not well described by the leading order \MADGRAPH\!+\PYTHIAE simulation. However, in the back-to-back region \HERWIGpp\ and \MADGRAPH\!+\cascade provide a better description of the shape of the jet multiplicity.
The measured differential cross section as a function of the transverse momentum of the four leading \pt jets is not  described by any of the LO predictions either in normalization or in shape. However,  \MADGRAPH\!+\cascade describes the shape of the distribution better than \MADGRAPH\!+\PYTHIAE .  

\begin{tolerant}{9600}
The predictions using dijet NLO matrix elements, \MG\!+\PYTHIAE(jj) and \MG\!+\cascade(jj) describe the lower multiplicity regions, as well as the transverse momenta of the leading jets, reasonably well.
The three-jet NLO calculation \MG\!+\cascade(jjj) describes very well the cross section of the third and fourth jets. 
\end{tolerant}

The measurements presented here provide stringent tests of theoretical predictions in the perturbative high-\pt and high-jet multiplicity regions.
Although the higher jet multiplicities are not described with either parton shower approach, it is interesting that 
the lower jet multiplicity cross section is described satisfactorily with NLO dijet calculations supplemented with \PBM -TMDs and TMD parton shower with fewer tunable parameters than in the case with conventional parton showers.

The measured observables and its statistical correlations are provided in HEPData~\cite{hepdata} as tabulated results.

\begin{acknowledgments}
	We congratulate our colleagues in the CERN accelerator departments for the excellent performance of the LHC and thank the technical and administrative staffs at CERN and at other CMS institutes for their contributions to the success of the CMS effort. In addition, we gratefully acknowledge the computing centers and personnel of the Worldwide LHC Computing Grid and other centers for delivering so effectively the computing infrastructure essential to our analyses. Finally, we acknowledge the enduring support for the construction and operation of the LHC, the CMS detector, and the supporting computing infrastructure provided by the following funding agencies: BMBWF and FWF (Austria); FNRS and FWO (Belgium); CNPq, CAPES, FAPERJ, FAPERGS, and FAPESP (Brazil); MES and BNSF (Bulgaria); CERN; CAS, MoST, and NSFC (China); MINCIENCIAS (Colombia); MSES and CSF (Croatia); RIF (Cyprus); SENESCYT (Ecuador); MoER, ERC PUT and ERDF (Estonia); Academy of Finland, MEC, and HIP (Finland); CEA and CNRS/IN2P3 (France); BMBF, DFG, and HGF (Germany); GSRI (Greece); NKFIH (Hungary); DAE and DST (India); IPM (Iran); SFI (Ireland); INFN (Italy); MSIP and NRF (Republic of Korea); MES (Latvia); LAS (Lithuania); MOE and UM (Malaysia); BUAP, CINVESTAV, CONACYT, LNS, SEP, and UASLP-FAI (Mexico); MOS (Montenegro); MBIE (New Zealand); PAEC (Pakistan); MES and NSC (Poland); FCT (Portugal); MESTD (Serbia); MCIN/AEI and PCTI (Spain); MOSTR (Sri Lanka); Swiss Funding Agencies (Switzerland); MST (Taipei); MHESI and NSTDA (Thailand); TUBITAK and TENMAK (Turkey); NASU (Ukraine); STFC (United Kingdom); DOE and NSF (USA).
	If acknowledgements for individuals are required for a short letter because some of the principal authors are funded through individual grants, it should be OK to add the lines below concerning individuals even for a short letter, but please first consult with the PubComm chair.
	
	\hyphenation{Rachada-pisek} Individuals have received support from the Marie-Curie program and the European Research Council and Horizon 2020 Grant, contract Nos.\ 675440, 724704, 752730, 758316, 765710, 824093, 884104, and COST Action CA16108 (European Union); the Leventis Foundation; the Alfred P.\ Sloan Foundation; the Alexander von Humboldt Foundation; the Belgian Federal Science Policy Office; the Fonds pour la Formation \`a la Recherche dans l'Industrie et dans l'Agriculture (FRIA-Belgium); the Agentschap voor Innovatie door Wetenschap en Technologie (IWT-Belgium); the F.R.S.-FNRS and FWO (Belgium) under the ``Excellence of Science -- EOS" -- be.h project n.\ 30820817; the Beijing Municipal Science \& Technology Commission, No. Z191100007219010; the Ministry of Education, Youth and Sports (MEYS) of the Czech Republic; the Hellenic Foundation for Research and Innovation (HFRI), Project Number 2288 (Greece); the Deutsche Forschungsgemeinschaft (DFG), under Germany's Excellence Strategy -- EXC 2121 ``Quantum Universe" -- 390833306, and under project number 400140256 - GRK2497; the Hungarian Academy of Sciences, the New National Excellence Program - \'UNKP, the NKFIH research grants K 124845, K 124850, K 128713, K 128786, K 129058, K 131991, K 133046, K 138136, K 143460, K 143477, 2020-2.2.1-ED-2021-00181, and TKP2021-NKTA-64 (Hungary); the Council of Science and Industrial Research, India; the Latvian Council of Science; the Ministry of Education and Science, project no. 2022/WK/14, and the National Science Center, contracts Opus 2021/41/B/ST2/01369 and 2021/43/B/ST2/01552 (Poland); the Funda\c{c}\~ao para a Ci\^encia e a Tecnologia, grant CEECIND/01334/2018 (Portugal); the National Priorities Research Program by Qatar National Research Fund; MCIN/AEI/10.13039/501100011033, ERDF ``a way of making Europe", and the Programa Estatal de Fomento de la Investigaci{\'o}n Cient{\'i}fica y T{\'e}cnica de Excelencia Mar\'{\i}a de Maeztu, grant MDM-2017-0765 and Programa Severo Ochoa del Principado de Asturias (Spain); the Chulalongkorn Academic into Its 2nd Century Project Advancement Project, and the National Science, Research and Innovation Fund via the Program Management Unit for Human Resources \& Institutional Development, Research and Innovation, grant B05F650021 (Thailand); the Kavli Foundation; the Nvidia Corporation; the SuperMicro Corporation; the Welch Foundation, contract C-1845; and the Weston Havens Foundation (USA).
\end{acknowledgments}
\bibliography{auto_generated}   
\cleardoublepage \appendix\section{The CMS Collaboration \label{app:collab}}\begin{sloppypar}\hyphenpenalty=5000\widowpenalty=500\clubpenalty=5000
\cmsinstitute{Yerevan Physics Institute, Yerevan, Armenia}
{\tolerance=6000
A.~Tumasyan\cmsAuthorMark{1}\cmsorcid{0009-0000-0684-6742}
\par}
\cmsinstitute{Institut f\"{u}r Hochenergiephysik, Vienna, Austria}
{\tolerance=6000
W.~Adam\cmsorcid{0000-0001-9099-4341}, J.W.~Andrejkovic, T.~Bergauer\cmsorcid{0000-0002-5786-0293}, S.~Chatterjee\cmsorcid{0000-0003-2660-0349}, K.~Damanakis\cmsorcid{0000-0001-5389-2872}, M.~Dragicevic\cmsorcid{0000-0003-1967-6783}, A.~Escalante~Del~Valle\cmsorcid{0000-0002-9702-6359}, R.~Fr\"{u}hwirth\cmsAuthorMark{2}\cmsorcid{0000-0002-0054-3369}, M.~Jeitler\cmsAuthorMark{2}\cmsorcid{0000-0002-5141-9560}, N.~Krammer\cmsorcid{0000-0002-0548-0985}, L.~Lechner\cmsorcid{0000-0002-3065-1141}, D.~Liko\cmsorcid{0000-0002-3380-473X}, I.~Mikulec\cmsorcid{0000-0003-0385-2746}, P.~Paulitsch, F.M.~Pitters, J.~Schieck\cmsAuthorMark{2}\cmsorcid{0000-0002-1058-8093}, R.~Sch\"{o}fbeck\cmsorcid{0000-0002-2332-8784}, D.~Schwarz\cmsorcid{0000-0002-3821-7331}, S.~Templ\cmsorcid{0000-0003-3137-5692}, W.~Waltenberger\cmsorcid{0000-0002-6215-7228}, C.-E.~Wulz\cmsAuthorMark{2}\cmsorcid{0000-0001-9226-5812}
% ...
\par}
\vskip\cmsinstskip
\dag:~Deceased\\
$^{1}$Also at Yerevan State University, Yerevan, Armenia\\
$^{2}$Also at TU Wien, Vienna, Austria\\
% ...
\end{sloppypar}
\end{document}